\documentclass[12pt]{article}
\usepackage{latexsym,amsfonts,epsfig,eucal, amsmath, amssymb,amsthm,tikz,pgfplots,stmaryrd}
\usepackage{enumitem}
\usepackage{tikz-cd}
\usepackage{tikz}
\usepackage{subcaption}
\usepackage{adjustbox}
\usetikzlibrary{arrows.meta, calc}
\usepackage{pgfplots}
\pgfplotsset{compat=1.15}
\usepackage{mathrsfs}
\usepackage[style=authoryear, backend=biber, natbib]{biblatex}
\usetikzlibrary{arrows}

\begin{document}
\title{Expected Utility from a Constructive Viewpoint}
\author{Kislaya Prasad\\University of Maryland\footnote{Address:
    Robert H. Smith School of Business, University of Maryland, College Park, MD
    20742, U.S.A. Telephone: +1 (301) 405-6359. Email: kprasad@umd.edu.}}
\date{February 13, 2023\\ This version: December 23, 2023} \maketitle
\begin{abstract}
This paper introduces a space of variable lotteries and proves a constructive version of the expected utility theorem. The word ``constructive'' is used here in two
senses. First, as in constructive mathematics, the logic underlying proofs is intuitionistic. In a second sense of the word, ``constructive'' is taken to mean ``built up from smaller
components.'' Lotteries as well as preferences vary continuously over some topological space. The topology encodes observability or verifiability restrictions -- the open sets of
the topology serve as the possible truth values of assertions about preference and reflect constraints on the ability to measure, deduce, or observe.
Replacing an open set by a covering of smaller open sets serves as a notion of refinement of information. Within this framework,
inability to compare arises as a phenomenon distinct from indifference and this gives rise to the constructive failure of the classical expected utility theorem. A constructive version of the theorem is then proved, and accomplishes several things. First, the representation theorem uses continuous real-valued functions as indicators of preference for variable lotteries and these functions reflect the inability to compare phenomenon. Second, conditions are provided whereby local representations of preference over open sets can be collated to provide a global representation. Third, the proofs are constructive and do not use the law of the excluded middle, which may not hold for variable lotteries. Fourth, a version of the classical theorem is
obtained by imposing a condition on the collection of open sets of the topology which has the effect of making the logic classical.

\bigskip


\smallskip


\smallskip

\noindent Declarations of interest: none
\end{abstract}
\newpage
\section{Introduction}
In classical expected utility theory, the decision-maker is assumed to have well-defined, pre-existing, preferences that are effortlessly interrogated. For any two
lotteries, it is always possible to assert either that one is better than the other, or that the decision-maker is indifferent between them. The possibility that the decision-maker
may not have information sufficient to make a determination of preference is effectively ruled out. A first goal of this paper is to allow for the possibility that assertions of
preference depend upon the stage of knowledge of the decision-maker, and that this knowledge depends upon some underlying process of measurement, computation, deduction,
or observation. The stage of knowledge is assumed to be incomplete and evolving, leaving open the possibility that in some stages there may be lotteries that
are not comparable. Indeed, there are assertions within mathematical systems, as well as about empirical phenomena, that cannot be finitely confirmed. In such
situations the available information will never be sufficient for making a determination of preference between pairs of lotteries whose value depends on the assertions. To this end,
the ``inability to compare'' phenomenon is not ruled out here and is systematically accounted for in the representation theorem. The paper abstracts from the specifics of the process of
observation or measurement so as to develop a completely general theory which can cover diverse concrete realizations of observation systems. Examples are used to
illustrate concrete cases.

A second goal of the paper is to develop a theory where assertions of preference between complex alternatives are pieced together from assertions about their parts. There
are several ways in which this kind of situation might arise. Different observations, measurements, or computations might lead to different assertions about preferences, and
the theory provides a way to combine such ``local'' assertions about preference, based on local information, into ``global'' assertions. In a similar vein, preferences between
high-dimensional alternatives may be constructed by first examining preferences between simpler low-dimensional alternatives, which are then combined to form solutions
for the original, more complicated, problem. The use of the word constructive in the title of the paper is used to suggest such a progression and ``building up'' of preferences
from its constituent parts. There is another well-established sense in which the word constructive is used in mathematics, which is to indicate that the logic underlying proofs is
intuitionistic. The classical expected utility theorem fails to be constructive in this sense. This paper provides a constructive version of classical expected utility theory and
proofs adhere to intuitionistic logic. A version of the classical theorem is obtained by imposing an assumption which has the effect of making the logic classical.

To address the above goals in a unified framework, this paper introduces the concept of a {\em variable lottery\/}. The space of variable lotteries comprises of continuous
functions from open subsets of $X$ to some space of lotteries (probability distributions over prizes or consequences). Preferences, which are defined over variable lotteries and
allowed to vary continuously, may be viewed as assertions attached to open subsets of $X$.\footnote{Importantly, preferences are {\em not\/} defined pointwise.}
The collection of open sets of $X$, denoted $\mathcal{O}(X)$, plays an important role here as the set of possible {\em truth values\/} of the system. In particular,
questions such as ``When is the variable lottery $p$ preferred to variable lottery $q$?'' or ``When is the decision-maker indifferent between variable lotteries $p$ and $q$?''
must be answered by selecting an open set in $\mathcal{O}(X)$. The classical setting is a special case of this framework where $X$ is the singleton space with open
sets $X = \{*\}$ and $\emptyset$. So, classically, the only possible answers to the above questions are $X$ (``always") or $\emptyset$ (``never"). However, when lotteries
and preferences vary over more general spaces, the answer is allowed to be some subset of $X$. The topology places restrictions on the possible answers. Consider,
for instance, two functions $f,g:\mathbb{R}\rightarrow \mathbb{R}$ with $f(x)=x$ and $g(x)=1$ and ask ``When are $f$ and $g$ equal?"
If open intervals in the standard topology are the only possible truth values, there is no non-empty open set over which $f=g$ (the set $\{1\}$ not being open).
In such cases, the truth value of $f=g$ is defined as the largest open set for which the assertion holds, {\em viz.\/} $\emptyset$.

The rationale for choosing open sets as truth values is discussed in greater detail in the Appendix but the general idea is to replace the classical
truth value of a statement with the conditions of verification of the statement. So, in the previous example, the question becomes ``When can $f$ and $g$
be {\em confirmed\/} to be equal?'' If the value of $x$ can only be asserted to lie within some interval $(a,b)$, however small, equality of $f$ and $g$ can never be
confirmed. One is back in a classical world if $\mathcal{O}(X)$ consists of {\em all\/} subsets of some finite set $X$. Here the requirement that possible answers have
open sets as truth values is not restrictive, since each $\{x\}$ is open. In this vein, a collection of open sets provides a model for the intuitionistic propositional calculus,
where truth coincides with verifiability. In contrast, a Boolean algebra provides a model for the classical propositional calculus (see \citet{goldblatt1982} and
\citet{maclane1996}). \citet{vickers1996} provides a detailed exposition of the link between topology and logic, emphasizing how open sets correspond to observable
properties.\footnote{His pithy slogan summarizes this point \citep{vickers1996}: ``Open sets are observable properties, and a point is what is being observed."
For a topology, arbitrary disjunctions and finite conjunctions of open sets are open, whereas the complement of an open set is not necessarily open. This accords with
the intuition that infinite conjunctions could not possibly be finitely confirmed, and that negation can change the observational content of a proposition.}
In the Appendix, a number of examples of observation, measurement, and computation are presented to support the claim that open sets capture the observational
content of statements. In light of the above, a useful way of reasoning about ranking lotteries begins with the assertion of an open set $U\subseteq X$ and
asking what rankings are entailed? For instance, if the truth value of ``$p$ is preferred to $q$'' contains $U$, then the assertion of $U$ entails the assertion
``$p$ is preferred to $q$''. Assertions of open sets can then be thought of as corresponding to observations, measurements, computations or deductions about
the point $x$ in $X$ that obtains (i.e., the assertion of $U$ is the confirmation that $x\in U$). In general, some assertion can be made at point $x\in X$ if there
is an open set $U$ which contains $x$, and the assertion is true for all points in $U$. $\mathcal{O}(X)$ can be chosen to impose bounds on the decision-maker's
ability to identify the stage exactly. Continuity also plays an important role here, capturing essential
computational and observational notions. The continuity of variable lotteries, and of preferences, ensures that observational constraints built into the topology are not violated.
One cannot, from measurements of the outcome of a mapping, make deductions or assertions about the stage that would not be possible using just the open sets.

An immediate consequence of the manner in which assertions of preference are made above is that, unlike in the classical setting,
there will be points in $X$ at which two variable lotteries are not comparable. This is exactly as in the comparison by equality of the two functions $f$ and $g$ above.
The following example illustrates this point -- suppose there are two variable lotteries $p$ and $q$
and a point $x\in X$ such that every open neighborhood $U$ of $x$ contains two open subsets $V_1$ and $V_2$ with $p$ preferred to $q$ for all $y\in V_1$ and $q$
preferred to $p$ for all $y\in V_2$. Then there can be no neighborhood of $x$ in which either one of $p$ or $q$ is preferred to the other. The decision-maker is also
not indifferent between $p$ and $q$.
In the case of variable lotteries, a distinction exists between indifference and non-comparability. Indifference can be defined in the usual way for a pair of variable lotteries as
the assertion that neither of the two is better than the other. However, the appropriate definition of negation in the current framework is as follows---for some proposition $P$,
we assert $\neg P$ over $U\in \mathcal{O}(X)$ if and only if $P$ does not hold for any open $V\subseteq U$. In the case of two lotteries $p$ and $q$, indifference over $U$
requires that for no $V \subseteq U$ can we assert that either of $p$ or $q$ is better than the other (when $p$ and $q$ are restricted to $V$). So, there are three possibilities:
the decision-maker (1)~is indifferent between $p$ and $q$, (2)~prefers one of $p$ or $q$ to the other, and (3)~is unable to compare $p$ and $q$.\footnote{It is important to
note that ``$p$ and $q$ are not comparable'' can never be an assertion of the system, because it may be proved false if future information allows the decision-maker to assert that
one is better than the other.} Since (1) and (2) are negations of one another, we have a violation of the law of the excluded middle, the principle of classical logic asserting that either a proposition is true or its negation is true. This will be seen to have significant implications for the development of the theory.

An appealing feature of the present approach is that it provides a natural way of thinking about refinement of information and of obtaining better approximations
of preference rankings. If one thinks of an open set $U$ as carrying information about the value of $x\in X$ that obtains, then
a smaller open set ($V \subset U$) provides more refined information. Lotteries that are not comparable over $U$ may become comparable over $V\subset U$.
Thus, refinement corresponds to replacing a set $U$ with a covering of smaller open sets $\{U_i\}$ with  $\bigcup_i U_i = U$.
It will be an assumption of the theory that if $U$ entails the assertion ``$p$ is preferred to $q$'', then this
ranking continues to hold when some $V \subset U$ is asserted (i.e., an assertion of preference, once made, cannot be retracted when more detailed information
becomes available). With this assumption, refinement preserves previously established rankings but also reveals new rankings, as variable lotteries that were previously
non-comparable become comparable.

The space of variable lotteries described above is a mathematical object called a sheaf \citep[see][]{maclane1996,rosiak2022sheaf}. This is a mapping $F$ that assigns a set $F(U)$ to each
$U\in \mathcal{O}(X)$, and a restriction map $\rho_{UV}:F(U)\rightarrow F(V)$ to each inclusion $V\subset U$, and satisfies certain properties. A formal definition of a
sheaf is included in the Appendix. Here a concrete example of a sheaf -- the set of all continuous real-valued functions defined over open subsets of $X$ -- will be used to illustrate its properties. First, each $f \in F(U)$ gets mapped by $\rho_{UV}$ to its restriction $f|_V$ in $F(V)$. Additionally, $\rho_{UU}=id$ and $\rho_{VW}\circ \rho_{UV}=\rho_{UW}$.
 The restriction of a continuous function to an open $V\subset U$ is clearly continuous, and the other properties are easily confirmed. The second is the ``gluing'' property for
 open covers (or coverings) of $U\in \mathcal{O}(X)$, i.e., a collection of open sets $\{U_i\}_{i\in I}$ such that $\bigcup_i U_i = U$.
 If $\{x_i| x_i\in F(U_i) \}$ is a collection of elements satisfying $x_i|_{U_i\cap U_j} =x_j|_{U_i\cap U_j}$ (i.e., they match on overlaps) then there is a
 unique $x\in F(U)$ such that $x|_{U_i} = x_i$ for all $i\in I$. In the case of continuous real-valued functions, if the functions $f_i:U_i \rightarrow \mathbb{R}$ are
 continuous for all $i\in I$, then there is at most one continuous function $f:U\rightarrow \mathbb{R}$ with $f|_{U_i}=f_i$ which exists if and only if the $f_i$
match on overlaps ($f_i(x) = f_j(x)$ for $x\in U_i\cap U_j$ for all $i,j\in I$). This second property is key to this paper's approach to the second goal of deducing assertions
about preferences between complex alternatives by piecing together assertions for simpler alternatives. An evocative motivation for sheaves is as ``probes'' of an unknown space
via mappings from known spaces (see \citet[][p.79]{rosiak2022sheaf} and the nLab entry \citep{Nlab}). In the present context, the set of variable lotteries $F(U)$ will be the
set of continuous functions from $U$ to a space of probability distributions. This may be interpreted as the set of probes from $U$ (a test space) to the space of probability
distributions. Assuming the conditions enumerated above for sheaves are satisfied, these probes give a complete and accurate picture of the space being probed.
By probing a complex space with simpler test spaces, it becomes possible to get an accurate picture of parts of the space into which the test space is mapped,
and the partial information can be pieced together to get the global picture.

There is a little more to say about some abstract ideas regarding sheaves to explain the mathematical intuition behind this paper and the strategy behind the
proofs.\footnote{This is not entirely necessary for an understanding of the results in the paper, but ensures completeness in attribution.}
The category of sheaves on a topological space $X$ may be regarded as a generalized universe of sets where all sets are varying continuously over $X$
(i.e., a universe of ``variable sets'').\footnote{\citet{maclane1996} develop the idea of a topos as a generalized universe of sets. The category of sheaves is a
topos, and can be interpreted as a universe of variable sets. \citet{lawvere2003sets} and \citet{lawvere2009conceptual} provide an exposition of this point of
view. One can specialize to the familiar universe of sets by taking $X$ to be the singleton space, or by
making all objects constant.} Within this mathematical universe, which will be denoted as Sh($X$), one can carry out familiar set-theoretic
constructions. For instance, one such construction is of the real numbers as a Dedekind cut of the rational numbers. This construction leads to the real numbers in
Sh($X$) being (isomorphic to) the sheaf of continuous real-valued functions over $X$ discussed above. Sh($X$) has an internal language within which one can
state formulae or sentences, just like in the familiar universe of sets. The important difference between the two is that the rules of inference for Sh($X$) are the rules
for the first order {\em intuitionistic\/} predicate calculus. This system of logic is like classical logic, except that the law of the excluded middle is absent. A consequence
is that theorems with constructive proofs can be carried over to the universe of sheaves, whereas classical results carry over only to some special cases (such as the singleton
space discussed earlier). Given the ``constructive flavor'' of the expected utility theorem,\footnote{Evident in the calibration at the center of the proof, but even more so in
expositions such as \citet{pratt1964}. } and the fact that the set of variable lotteries and the set of all continuous real-valued functions are sheaves, it is a reasonable
conjecture that one can state and prove the expected utility theorem for Sh($X$). This would require defining suitable preference orderings over variable lotteries, and
finding an order-preserving mapping from variable lotteries to continuous real-valued functions. It should be possible to keep the statement of the expected utility theorem
unchanged and to prove this result using suitable modifications of the axioms of classical expected utility theory. The familiar weak order,
independence, and continuity assumptions do need modification, and sheaf semantics provides a guide for carrying the axioms over to Sh($X$). Such a theorem
would achieve the first goal of the paper. In Sh($X$) real-valued functions $r, s: X\rightarrow \mathbb{R}$ are ordered by the ``less than'' relation ($<$) in the same
way that lotteries are ranked: at some $x\in X$, $r<s$ if and only if there is an open neighborhood $U$ of $x$ such that for all $y\in U$ $r(y)<s(y)$ (equivalently,
the truth value of $r<s$ is the largest open set over which $r<s$). So, one can seek a representation where if  ``$p$ is preferred to $q$'' can be asserted then
it is also possible to assert that the expected utility of $p$ is greater than the expected utility of $q$. Similarly, indifference between $p$ and $q$ would be
expected to entail equality of their expected utilities. If $p$ and $q$ were not comparable over $U$ in terms of preference,
then the expected utilities of $p$ and $q$ would not be comparable in terms of $<$ over $U$ either.

Unfortunately, the expected utility theorem turns out to be non-constructive. This is shown in Example 2, where preferences are constructed to satisfy the three axioms,
but a representation using continuous real-valued expected utility functions is not possible.
The difficulty relates to stages in $X$ where no two lotteries are comparable. In the classical setting, if no two lotteries can be ranked in terms of strict preference
the decision-maker is judged to be indifferent between every pair of lotteries (by an application of the law of
the excluded middle). This cannot be done in the constructive setting. One approach for obtaining the intended result is to narrow down
the class of preferences. This is done via an additional assumption called minimal comparability. On {\em a priori\/} grounds, it is possible to assert that either
(1)~there exist at least two lotteries $p$ and $q$ such that one is preferred to the other, or (2)~the decision-maker is indifferent between all lotteries. Together with
the standard assumptions, this condition is sufficient to ensure representation of preferences via continuous real-valued expected utility functions. Specifically,
for every $W\in \mathcal{O}(X)$ there is a mapping $u_W$ from $F(W)$ to the set of continuous real-valued functions with domain $W$
which send every variable lottery to its expected utility. Further, for every $V\subset W$, there is a mapping $\rho_{WV}: F(W)\rightarrow F(V)$, such that
$\rho^\prime_{WV} \circ u_W =u_V \circ \rho_{WV} $ (here $\rho^\prime_{WV}$ is the restriction mapping for continuous real-valued functions).
Such a $u$ is called a natural transformation. This result is Theorem~1 of the paper.
However, the necessity part of the classical expected utility theorem no longer holds. The minimal comparability requirement
is {\em local\/} in the sense that one of the above two conditions is assumed to hold in some open neighborhood of every point $x\in X$.
However, it could be that condition~(1) is satisfied for a different pair of lotteries in different
parts of $X$. Then we would have an expected utility representation for every lottery in an open neighborhood of each stage in $X$, but it may not be possible to ``glue'' them together
into a global function over all of $X$. A slightly stronger version of the minimal comparability condition is needed to ensure a representation using a {\em global\/} utility
function. This result is proved in Corollary~1. In Theorem~1 and Corollary~1, both lotteries and preferences are allowed to vary. For Corollary~2, the requirement is imposed
that all constant lotteries are comparable and their rankings never change over $X$---i.e., for any pair of constant lotteries $p$ and $q$, either $p$ is preferred to $q$ over all of $X$
or $p$ is never preferred to $q$ (and similarly for $p$ indifferent to $q$ and $q$ is preferred to $p$). This assumption
has the effect of arresting the variation in the utility function so that the utility of any consequence or prize (i.e., the utility of the degenerate distribution on the prize) becomes constant
over $X$. The expected utility function is still variable, but all the variation comes from the variation in probabilities.

The second approach to obtaining a representation result is by assuming that the sheaf satisfies the law of the excluded middle---for all propositions $P$ either $P$ or $\neg P$. An equivalent formulation of this classical logic assumption is the following requirement for the topology $\mathcal{O}(X)$:
for every open set $W$ in $\mathcal{O}(X)$, $W \cup \mbox{Int}(W^c) = X$, where $\mbox{Int}(W^c)$ is the interior of the
complement of $W$. This law is, of course, implicitly assumed in the classical expected utility theorem. A version of the classical result is presented in Theorem~2. Together with the weak order, independence, and continuity axioms, the classical logic assumption provides necessary and sufficient conditions for the representation of preferences using a continuous real-valued expected utility function. As might be expected, the phenomenon of non-comparability now disappears.

Finally, it is worth observing that the proofs of Theorem~1 and Corollary~1 are constructive. There is a long tradition in mathematics which
insists that objects proved to exist should be algorithmically realizable or computable \citep[see][]{bridgesrichman1987}.\footnote{The appendix of this paper
also provides a brief survey.}
This tradition rejects use of the law of the excluded middle, which is the principle of classical logic stating that for every proposition $P$, either $P$ is true or its
negation $\neg P$ is true. Constructivists also reject proof by contradiction (for every $P$, $\neg\neg P \Rightarrow P$) since this is an implication of the law of the
excluded middle. These logical rules are the culprits
behind the fact that using classical logic it is possible to prove existence of objects that are not algorithmically realizable.
The logical system used in constructive mathematics, distinguished principally by the omission of the law of the excluded middle, is intuitionistic logic. The current paper
is not concerned with the philosophical debates about how mathematics ought to be conducted. However, a very practical issue arises in the framework outlined
above---the law of excluded middle does not hold. There was an intimation of this in the demonstration above that comparisons of a pair of lotteries lead to
three distinct possibilities: (1) one lottery is judged to be better than the other, (2) the decision-maker arrives at the conclusion that she is indifferent between the two, or
(3) she is unable to make a comparison. For this reason, only constructive arguments were used when proving results.

{\it Related Literature.\/} The word ``constructive'' has been used in a different sense in the recent decision theory literature, but the goals are related. A central idea is to begin with a decision problem confronting a decision-maker and then to characterize the process by which value judgments and probability assessments are formed to guide choices. In this vein, \citet{blume2021} begin with a natural language description of a decision problem and derive states, outcomes, and Savage acts, and from this derive an expected utility representation. \citet{shafer2016} makes the point that a fully constructive theory should include an account of how probability and value judgments are formed, and then inform preferences. In particular, in his view, preferences should follow probabilities and utilities instead of the latter being elicited from preference. He is sceptical of the ability of expected utility to provide such a constructive theory, and presents an alternative. Perhaps the closest to the present paper is \citet{kaneko2020}, who models a form of bounded rationality in which permissible probabilities are restricted to decimal fractions of finite depth up to a given bound. This leads to the phenomenon of non-comparability, with the problem being worse when
the cognitive bound is smaller. He obtains a representation in terms of a two-dimensional vector-valued utility function with the interval order. The expected utility hypothesis holds in his framework when the bound goes to infinity. One sense in which the term ``constructive'' is used in the present paper is taken from constructive mathematics. In this sense of the word, the constructive development of preference and utility was pioneered by Douglas Bridges \citep{bridges1982, bridges1989,bridges1994}) within the constructive framework of Errett Bishop \citep{bishop1967}. This is a framework that does not allow use of the law of the excluded middle). \citet{richter1999} present a development within computability theory. The present paper focuses on choice under uncertainty and models preference and expected utility in a setting with variable lotteries, with a topology being used to encode essential observational or computational restrictions (in other words, within the category of sheaves). The extensive literature upon which this theory is built is discussed in the appendix of the paper.

The following is an outline of the remainder of the paper. The appendix and the end provides a short survey of key ideas in constructive mathematics and sheaf theory with the objective of making the paper relatively self-contained. It also presents a number of examples of phenomena involving measurement, computation, and observation, and illustrates how an intuitionistic logic applies, and how these can be put into a common topological setting. Section~2 contains the main results paper, including the constructive expected utility theorem for variable lotteries, and its specialization when classical logic applies. Section~3 contains concluding remarks. The proofs are at the end.

\section{The Expected Utility Theorems}
In this section, the status of the expected utility theorem is examined within the framework of variable lotteries described in the introduction. Variable lotteries are formally defined in section 2.1, where it is shown that the logic of assertions about variable lotteries is constructive. The standard axioms -- weak ordering of preferences, independence, and continuity -- are suitably generalized for the framework, after which a version of the theorem appropriate for the setting of variable lotteries can be formulated. However, the classical theorem fails in this setting, with the failure being a direct consequence of the fact that the classical expected utility theorem is not constructive. This constructive failure is demonstrated in section~2.2. After this, two main theorems are proved. A constructive expected utility theorem is proved in section 2.3. Then, in section 2.4, using an assumption that ensures that the law of the excluded middle holds, a classical version of
the expected utility theorem is shown to hold.

\subsection{The Framework of Variable Lotteries}
Let $X$ be a topological space equipped with an open set collection $\mathcal{O}(X)$. $X$ is the space over which lotteries vary, and  $\mathcal{O}(X)$ is the set of possible truth values of assertions within the framework. In the classical framework there are only two truth values, {\bf true\/} and {\bf false\/}. Here any $U \in \mathcal{O}(X)$ can be chosen to specify {\em when\/} a proposition holds. The restriction to open sets, instead of all subsets of $X$, encodes the condition of truth as verifiability as discussed in the introduction and Appendix~1.
On occasion, it will also be useful to view $\mathcal{O}(X)$ as a poset ordered by inclusion.
The consequences, or prizes, will be denoted by ${\bf Z}(X)$, a finite set of locally constant functions defined on $X$. Let $n$ be the number of elements of ${\bf Z}(X)$. Elements of ${\bf Z}$ are enumerated as $\{z_1, \ldots, z_n\}$.\footnote{The elements of {\bf Z} could have been  allowed to vary over $X$. However, this is not assumed so as to more clearly illustrate the effect of  variation in preferences.}
${\bf Z}(U)$ denotes the set whose elements are the restriction of these functions to the open set $U \in \mathcal{O}(X)$.\footnote{It is possible to allow new elements to appear upon restriction (e.g., functions that are defined only over some open $U \subset X$), although this is not done in the case of ${\bf Z}$. The important requirement is that an element $f$ of $Z(X)$ continues to exist as $f|_U$ in $Z(U)$. However, preferences will be permitted later to have only a partial domain of definition, and the present simplification permits a clearer articulation of that phenomenon.} The set of rational numbers is denoted by $\mathbb{Q}$, and the classical real number line is denoted $\mathbb{R}$. Let ${\bf Q}(X)$ denote the set of locally constant functions $\{q: X \rightarrow \mathbb{Q}\}$. This includes all constant rational functions on $X$, but possibly also non-constant functions when the domain is not connected (for instance, if $X = (-1,0)\cup (0,1)$, we could have $q(x)=1$ for $x<0$, and $q(x) = 2$ for $x>0$).  ${\bf R}(X)$ will denote the set of {\em continuous\/} functions $\{r: X \rightarrow \mathbb{R}\}$. ${\bf Q}(U)$ and ${\bf R}(U)$ will denote, respectively, the locally constant rational and continuous real-valued functions with domain $U \in \mathcal{O}(X)$. In the case of ${\bf R}(U)$ new functions may appear (i.e., functions that were not present in ${\bf R}(X)$). This is because there may exist functions in ${\bf R}(U)$ that cannot be extended to continuous real-valued functions over
$X$ (e.g., $f(x) = {1 \over x}$ exists in ${\bf R}((0,1))$ but not in ${\bf R}((-1,1))$).

Probabilities  on ${\bf Z}$ are defined next. Let $\mathbb{S}^n \equiv \{ (p_1, p_2, \ldots, p_n) \in \mathbb{R}^n_+ | \sum_{i=1}^n p_i = 1\}$. Elements of $\mathbb{S}^n$ will also be allowed to vary over $X$. This yields the {\em variable lotteries\/}:
$$ S(X) = \{f: X \rightarrow \mathbb{S}^n| f \mbox{ \ is continuous}\}$$
with the $S(U)$ defined analogously for each $U \in \mathcal{O}(X)$. So, $p_i(x)$ is the probability of $z_i(x)$. Let $\delta_i \in S(X)$ denote the function that always assigns probability one to $z_i$ (and probability 0 to all $z_j$ with $j\neq i$).
From this, any $f\in S(X)$ can be written as $f (x)= \sum_{i=1}^n p_i(x)\delta_i(x)$ for all $x\in X$.
For $f \in S(X)$, the notation $f|_U$ is used to identify the corresponding restriction of $f$ in $S(U)$.
For $V \subset U$, the mapping $\rho_{UV}: S(U)\rightarrow S(V)$ which sends each $p\in S(U)$ to its restriction
$p|_V$ in $S(V)$ is called the restriction mapping. As with ${\bf R}$, new variable lotteries may arise upon restriction to $U \subset X$.  Observe that ${\bf R}(X)$, ${\bf Q}(X)$, and $S(X)$ are sheaves, and satisfy the conditions enumerated in the definition of a sheaf in Appendix~1. For ${\bf R}(X)$ and $S(X)$ above, pointwise sum, product and scalar multiplication are defined in the familiar way. In the case of $S(X)$, the natural operation derived from these is the weighted average defined as follows. For $p,q \in S(X)$ and given a continuous function $a:X \rightarrow [0,1]$, the continuous function $ap + ({1}-a)q \in S(X)$
is defined by: \[ a(x) p_i(x) + (1-a(x))q_i(x) \quad \mbox{for each {\em i\/} $\in \{1, \ldots, n\}$ and all $x\in X$. } \]

Preferences are defined over $S(X)$ and also over $S(U)$ for each $U\in \mathcal{O}(X)$. For $U\subseteq X$, the preference relation over $S(U)$ may be described as a subset of $S(U) \times S(U)$ in the usual manner.
Using the standard notation, $p \prec q$ will be used to denote ``$p$ is worse than $q$'', or ``$q$ is preferred to $p$''.
An important difference is that preferences are not defined pointwise but for open sets $U\in \mathcal{O}(X)$.
The assertion $p \prec q$ may hold only over some part of $X$ and, since the truth value of assertions are required to be open sets,
it will only be possible to assert $p \prec q$ over some $U \in \mathcal{O}(X)$. $U$ identifies the part of $X$ where the assertion holds.\footnote{Informally, the theory involves rankings of ``shreds" of functions.} Preferences over the sets $S(U)$ for different $U\in \mathcal{O}(X)$ are not arbitrary, but will be related in a manner to be detailed later.
In light of this, it will make sense to speak of preferences as being defined over the whole structure $S$.

For an open set $U\in \mathcal{O}(X)$, and $p,q: U \rightarrow \mathbb{S}^n$, the ``forcing notation'' is used to write $U \Vdash p \prec q$ when the assertion $p \prec q$
is entailed by $U$, or holds over all of $U$.\footnote{See Appendix~1 for a more precise definition of forcing.}
The assertion of preference at a point $x$ can be derived from preference defined for open sets. The assertion $p(x)\prec q(x)$ will mean
that $x$ has an open neighborhood (i.e., some $U\in \mathcal{O}(X)$ with $x\in U$) such that the assertion is true for all points in the neighborhood. $U \Vdash p \prec q$ is read as
``$U$ forces $p \prec q$.'' Negation of assertions about preferences, and the relations $\precsim$ (``at least as good as") and $\sim$ (``indifference") can now be defined:
\begin{itemize}
\item $U \Vdash \neg(p \prec q)$ if and only if for no non-empty $V\subseteq U$ is it the case that $V \Vdash p \prec q.$
\item $U \Vdash p \sim q$ if and only if  $U \Vdash \neg(p \prec q)$ and $U \Vdash \neg(q \prec p).$
\item $U \Vdash p \precsim q$ if and only if $U \Vdash \neg(q \prec p)$.
\end{itemize}
Note that $p \precsim q$ is {\em not\/} equivalent to ``$p \prec q$ or $p\sim q$".
As a direct consequence of the fact that truth values can only be open sets, two variable lotteries $p$ and $q$ may sometimes not be comparable at some point $x\in U$ even if the domain of definition of their preference is $U$. The general point can be made in the context of continuous real-valued functions assuming the standard topology on $\mathbb{R}$. Two functions, $f$ and $g$, are depicted in Figure~1.
The function $f$ is defined to be equal to 0 for $x<0$, and equal to $x^2$ for $x\geq 0$. Function $g$ is defined to be equal to $0.5$. At the origin, consider if it is possible to assert $f=0$? This cannot be asserted because there is no open set $U \subseteq X$ containing 0 for which $f(x)=0$ for all $x\in U$. Is it possible to assert $f\neq 0$? Again no, because every open interval containing 0 contains open intervals where $f=0$. It is possible to assert $f=0$ for any $U \subseteq (-\infty, 0)$ and one may assert $f \geq 0$ (i.e. for all $U \subseteq (-\infty, +\infty)$). Equality of $f$ and $g$ can never be asserted. Any open interval that contains the $\hat{x}$ (the value for which $f$ and $g$ intersect) always contains open subintervals where $f<g$ and open subintervals where $f>g$. These properties carry over to variable lotteries and preferences defined on them.\footnote{Indeed, one can take ${\bf Z}= \{z_1, z_2\}$ so that a variable lottery can be represented by one real-valued function denoting the probability of $z_1$. If $z_1$ is always the better prize, then restricting $X$ to be $[-1,1]$ (to ensure values in $[0,1]$) allows the above example to be directly carried over to lotteries.} Consequently, completeness of $\precsim$ on $S(X)$ cannot be asserted.\footnote{Note that in an alternative development of expected utility theory with $\precsim$ as a primitive, the assumption that $\precsim$ is complete and transitive serves as the definition of a weak ordering.} Completeness of $\precsim$ can be derived from the weak order assumption if the law of the excluded middle holds. For continuous real-valued functions, if the law of the excluded middle were valid, $f<g$ and $\neg (f<g)$ would exhaust the set of possibilities. In Figure~1, there is the third possibility that $f$ and $g$ are not comparable. However, ``$f$ and $g$ are not comparable'' cannot be an assertion of the system, because such an assertion could become false.

\begin{figure}
\begin{center}
\begin{tikzpicture}
  \draw[->] (-2, 0) -- (3, 0) node[right] {$x$};
  \draw[->] (0, -1) -- (0, 3) node[above] {$y$};
  \draw[line width=0.3mm][scale=0.5, domain=0:2.4, smooth, variable=\x, black] plot ({\x}, {\x*\x});
  \draw[line width=0.3mm][scale=0.5, domain=-5:0, smooth, variable=\x, black] plot ({\x}, {0});
  \draw[line width=0.3mm][scale=0.5, domain=-5:6, smooth, variable=\y, black]  plot ({\y}, {1});
  \node [color=black] at (0.85,2.5) {$f$};
  \node [color=black] at (2.5,0.75) {$g$};
  \draw [dashed] (0.5,0) -- (0.5,0.5)node[color=black] at (0.5, -0.25) {$\hat{x}$};
\end{tikzpicture}
\caption{Ordering continuous functions}
\end{center}
\end{figure}

Theorem~1 below represents preferences over variable lotteries using continuous real-valued functions.
So, $u(f)$ is a function in ${\bf R}(X)$ such that, at each $x\in X$, $u(f)(x)$ is the utility of variable lottery $f$ at $x$.
The mapping $u$ from variable lotteries over all $W\subseteq X$ to continuous real-valued functions in ${\bf R}(W)$
is a more complicated object and will be discussed below.\footnote{It is a natural transformation of sheaves.\/}
One complication that will be shown to arise is that, for some $f \in S(X)$, there may be no continuous real-valued indicator
of preference over all of $X$. However, an indicator may exist for $U \subset X$. Alternatively stated, the continuous utility function of $f$ over $U$ cannot be extended to a continuous function over all of $X$. In this kind of situation it turns out to be the case that the assumptions enumerated below as requirements for Theorem~1 can be collectively satisfied over $U$, but not over $X$. Restricting rankings involving $f$ to $U$ (i.e., limiting the domain of definition of preference for $f$) then permits a representation for $f$ over $U$.\footnote{The inevitability of having to work with partially defined preferences will become apparent in Example~2 below which shows the constructive failure of the classical expected utility theorem. The domain of definition of preference for some variable lottery $f$ may also be thought of as its {\em type\/}. Only lotteries of the same type are then compared. Eventually, the expected utility of such an $f$ will be a partial function and will have the same type.} Proceeding with preferences defined on $S(X)$ will result in excluding situations of (a)~only partially defined numerical representations of preferences, and (b)~representations of preferences for a subspace of variable lotteries excluding the $f\in S(X)$ which do not have a continuous numerical representation. To account for such situations preferences will be defined on a structure closely related to $S(X)$. This new sheaf $P$ is defined as follows. For each $W \subseteq X$,
$$ P(W) = \left\{ f: W \rightarrow \sum_{i\in \Phi(W)} p_i\delta_i \middle|   \ p_i:W \rightarrow [0,1], \sum_{i\in \Phi(W)} p_i = 1,
p_i \ \mbox{continuous}  \right\},  $$
where $\Phi(W)$ is some subset of $\{1, 2, \ldots, n\}$. This is different from $S(W)$ because it fixes the support of the variable lotteries
over $W$ to a subset of $\Phi(W)$. If $\Phi(W) = \{1, 2, \ldots, n\}$, $P(W)=S(W)$. As before, $\delta_i\in S(W)$, so has $n$ coordinates with the $i$-th coordinate set to $1$ while the rest are $0$.  Implicitly, $p_j=0 \ \mbox{for} \  j\not\in \ \Phi(W)$. $P(W)$ is a subspace
of $S(W)$. It continues to be the case that if $f, g \in P(W)$ then, for any continuous $a:W \rightarrow [0,1]$, $af+(1-a)g \in P(W)$.
For $V \subseteq U$, the restriction mapping $\rho_{UV}: P(U) \rightarrow P(V)$ sends every $f \in P(U)$ to its restriction $f|_V \in P(V)$. $P(V)$ may however contain elements that are not restrictions of elements of $P(U)$. Lotteries with for which preferences are defined over $W$, continue to have preferences defined for $V\subset W$, so that $\Phi(W) \subseteq \Phi(V)$.
Once preferences $\prec$ are defined on $P$ then the requirement that they satisfy the assumptions of Theorem~1 enumerated below necessarily excludes from $P(W)$ any variable lotteries whose domain of definition of preferences does not include $W$.

The assumptions for preferences over $P$ are now stated. The first is a modified version of the classical weak order assumption, presented below as Assumption~1. The first part of Assumption~1 restates asymmetry and negative transitivity of $\prec$ for the case of functions. The second and third part of the definition are new. If $U \Vdash p \prec q$, it is reasonable to require that for all $V \subseteq U$, $p|_V \prec q|_V$. In other words, that the restriction operation is order-preserving. This is the case, for instance, when replacing an open set by one of its open subsets corresponds to refinement of information.
If coarser information is sufficient to deduce $p \prec q$, then this should also hold for the refinement. An explicit example is provided in Figure A1 of Appendix~1, where this property appears as the idea of ``persistence of truth in time.'' The order-preserving property of preferences with restriction is presented below as the second part of Assumption~1. The third part of Assumption~1 specifies the conditions under which it is possible to assert $U \Vdash p \prec q$  based on such assertions on parts of $U$.
This relies on the notion of {\em compatible\/} or {\em matching\/} families. Suppose we are given an open cover $\{U_i\}_{i\in I}$ with $\bigcup_{i\in I} U_i = U$. Then the set of variable lotteries $\{p_i:U_i \rightarrow \mathbb{S}^n\}$ is said to form a compatible family if for all $i$ and $j$ in $I$, $p_i = p_j$ for all $x\in {U_i\cap U_j}$. In this case, since $S$ is a sheaf, it follows that there is a unique $p:U \rightarrow \mathbb{S}^n$ such that $p_i = p|_{U_i}$ for all $i$. The compatible family will be said to collate to $p$.
The assumption states that for two compatible families $\{ p_i\}$ and $\{q_i\}$ that collate to $p$ and $q$ respectively, if $q_i$ is preferred to $p_i$ over each $U_i$, then $q$ is preferred to $p$ over all of $U$. The latter two properties of Assumption~1 are instances of general sheaf properties called {\em monotonicity\/} and {\em local character\/} respectively \citep[][p.~316]{maclane1996}. The need for the third property will be evident from the following example.\footnote{It was argued earlier that two variable lotteries $p$ and $q$ are uncomparable over $W$ if $W$ has subsets $W_1$ and $W_2$ with $W_1 \Vdash p \prec q$ and $W_2 \Vdash q \prec p$. This property rules out some other avenues for lotteries being uncomparable.}

\bigskip

\noindent {\bf Example 1.\/} Let $X = [0,1]$ and define a preference ordering on $P(U) = \{{p}:U\rightarrow \mathbb{S}^2 \}$ as follows. For each {\em proper\/} open subset $U \subset X$, ${p}, {q} \in P(U)$ are ordered using the standard inequality $<$ on real numbers as follows. Denoting ${p}$ by $(p_1, p_2)$ and ${q}$ by $(q_1, q_2)$, $U \Vdash {p} \prec {q}$ if and only if $p_1(x)<q_1(x)$ for every $x\in U$. Define every distinct pair of functions in $P(X)$ as being not comparable (for all ${p}$ in $P(X)$ it is still the case that ${p} \sim {p}$). For an open covering $\bigcup_i U_i$ of $X$ consider compatible families $\{{p}_i:U_i\rightarrow \mathbb{S}^2 \}$ and $\{{q}_i:U_i\rightarrow \mathbb{S}^2\}$ collating to ${p}$ and ${q}$ respectively. Suppose $U_i \Vdash {p}_i \prec {q}_i$ for all $i$. The preferences defined here could not possibly be represented by continuous real-valued functions ordered by $<$, because $({\bf R},<)$ satisfies property~3 of Assumption~1. $\blacksquare$

\bigskip

The formal definition follows. The name {\em weakly ordered sheaf\/} will be used to describe the object below.
\newtheorem{axiom}{Assumption}
\begin{axiom}[Weak Order] Preferences satisfy the following conditions defining a weakly ordered sheaf.
\begin{enumerate}
\item For every $W \in \mathcal{O}(X), P(W)$ is a weakly ordered set. In other words, preferences are (locally) asymmetric and negatively transitive:
\begin{enumerate}
 \item For all $p, q \in P(W)$, $W \Vdash p \prec q \Rightarrow \neg(q \prec p).$
\item For all $p, q, r \in P(W)$, if $W \Vdash p \prec q$ then there is an open cover $\{ W_i\}$ with $\bigcup W_i = W$ such that for each index $i$ either $W_i \Vdash r|_{W_i} \prec q|_{W_i}$ or $W_i \Vdash p|_{W_i} \prec r|_{W_i}$.
\end{enumerate}
\item $W \Vdash p \prec q$ if and only if for all open $V \subseteq W$, $V \Vdash p|_V \prec q|_V$.
\item Suppose we are given an open cover $\{W_i\}$ with $\bigcup W_i = W$, and $\{p_i:W_i \rightarrow \mathbb{S}^n\}$ and $\{q_i:W_i \rightarrow \mathbb{S}^n\}$ are two compatible families that collate to $p:W \rightarrow \mathbb{S}^n$ and $q:W \rightarrow \mathbb{S}^n$  respectively. If $W_i \Vdash p_i \prec q_i$ for all $i$ then $W \Vdash p \prec q$.
\end{enumerate}
\end{axiom}
The sheaf of continuous real-valued functions with the standard ordering (${\bf R}$, $<$) is a weakly ordered sheaf.
Figure~2 illustrates the need for the modified negative transitivity definition. Three continuous real functions, $p, q,$ and $r$, are depicted there (with domain $\mathbb{R}$ and the standard topology). Consider their ordering by $<$. Clearly, $p<q$ over the interval $(0,w)$. Negative transitivity would appear to fail because $r$ does not lie either below $q$ or above $p$ over this interval. However, one can find an open covering $\{W_i\}$ of $(0,w)$ such that for each index $i$ either $W_i \Vdash r|_{W_i} < q|_{W_i}$ or $W_i \Vdash p|_{W_i} < r|_{W_i}$. So the condition holds in a local sense. This property is assumed for the ordering $\prec$ on $P(X)$.

\begin{figure}
\begin{center}
\begin{tikzpicture}
  \draw[->] (-0.2, 0) -- (4, 0) node[right] {$x$};
  \draw[->] (0, -2.2) -- (0, 4) node[above] {$y$};
  \draw[line width=0.3mm][scale=0.5, domain=0:6, smooth, variable=\x, black] plot ({\x}, {0.15*\x*\x});
    \draw[line width=0.3mm][scale=0.5, domain=0:6, smooth, variable=\x, black] plot ({\x}, {1.5+0.15*\x*\x});
 \draw[line width=0.3mm][scale=0.5, domain=0:6, smooth, variable=\y, black]  plot ({\y}, {-4+2*\y});
 \draw[line width=0.1mm][scale=0.5, domain=0:8, smooth, variable=\y, black]  plot ({5.5}, {\y});
  \node [color=black] at (0.5,1) {$q$};
  \node [color=black] at (2.5,1.5) {$p$};
  \node [color=black] at (2.5,3.5) {$r$};
    \node [color=black] at (2.8,-0.25) {$w$};
\end{tikzpicture}
\caption{Negative transitivity for continuous functions}
\end{center}
\end{figure}

The next two familiar axioms are modified for the context of functions.
\begin{axiom}[Independence] For all $W\subseteq X$, $p, q, r \in P(W)$, and given a continuous function $a: W \rightarrow (0, 1]$,
$$ W \Vdash p \prec q \Rightarrow ap+({1}-a)r \prec aq+({1}-a)r.$$
\end{axiom}

\begin{axiom}[Continuity] For all $W\subseteq X$ and $p, q, r \in P(W)$, if $p \prec q \prec r$ over $W$
then there exists an open cover $\{W_i\}$ of $W$ such that,
for each $i$, there are continuous functions $a_i, b_i: W_i \rightarrow (0,1)$ with
\begin{quote}
$W_i \Vdash q \prec a_ip + ({1}-a_i)r$ and
$W_i \Vdash b_ip + ({1}-b_i)r \prec q$.
\end{quote}
The functions $a_i$ and $b_i$ can be chosen to take values in $\mathbb{Q}$ (i.e., $a_i, b_i: W_i \rightarrow (0,1)\cup \mathbb{Q}$).
\end{axiom}

In the continuity assumption, the consequent is asserted to be true for every element of some open cover of $W$. This would be implied if the existence of a global
continuous function was assumed directly. However, the weaker statement suffices, and is consistent with how existential quantifiers are treated in sheaves. Additionally, the $a_i$ and $b_i$ can be chosen to be rational-valued functions, which simplifies some proofs. The replacement of $a_i$ and $b_i$ with rational functions relies on the order-denseness of the rational numbers in the real numbers. This is classically true, but the corresponding result need not hold for two continuous real-valued functions over the entire domain $X$ -- it is not necessarily the case that there is a $w \in {\bf Q}(X)$ between them. However, there always exists an open cover $\{U_i\}$ of $X$ and $w_i \in {\bf Q}(U_i)$ such that each $w_i$ lies between the two functions. Note also the continuity of $a_i$ and $b_i$, which plays an important role in the theorem.

Recall that the definition of $P$ was motivated by the need to account for lotteries that are not representable by continuous real-valued functions on some part of $X$. In the case of the axioms above, the question arises whether the axioms still make sense if for some $p\in P(W)$ preferences are not representable. With Assumptions 1--3, the only lotteries that can be included in $P(W)$ are those that are representable over $W$. Assumption~4 below will ensure that each $P(W)$ is non-empty. The discussion after Theorem~1 below provides an explicit example of failure of the axioms to hold when preferences are not representable over $W$.

\subsection{Constructive Failure of the Expected Utility Theorem}
In expected utility theory the weak order, independence, and continuity axioms are necessary and sufficient to yield the representation of preferences by an expected utility function.
With the three modified axioms, one may conjecture that there is a corresponding representation theorem for variable lotteries.\footnote{The basis for this conjecture is discussed in the introduction and, in greater detail, in the appendix.\/}
Is there a continuous function $f$, possibly itself varying over $X$, such that $U \Vdash p \prec q$ if and only if $U \Vdash f(p) < f(q)$? Unfortunately, the classical theorem is non-constructive, and the difficulties relate directly to the non-constructive steps in the proof. The next example illustrates that such a representation may not exist. In this example $X=\{0, 1, 2\}$ denotes stages of mathematical knowledge. There is the following partial ordering of the stages:  $0\sqsubseteq 1$, $0 \sqsubseteq 2$, and for all $x \in \{0, 1, 2\}$, $x \sqsubseteq x$. Stages $1$ and $2$ are said to be {\it more refined\/} than $0$, in the sense that they include all the mathematical assertions that can be made at 0, and more. Stages $1$ and $2$ are not similarly comparable by this ordering because they represent alternative future paths of knowledge (for instance, a mathematical conjecture is proved to be true at $1$, false at $2$, and unresolved at $0$). A topology on $X$ is defined by the collection of upper sets of this poset (the Alexandrov topology). This topology is discussed in greater detail in Appendix 1 (see, in particular, the corresponding Figure~A1). The example is discussed by \citet[][pp. 28-29]{vickers1996} and \citet[][p. 189]{goldblatt1982}, among others. Vickers places the example in terms of computation, and the state 0 represents a computation that will never finish. An ordering of stages of knowledge of this kind is the basis for Kripke's semantics for intuitionistic logic \citet{kripke1965}.

 \bigskip

\noindent {\bf Example 2 (Constructive Failure of the Expected Utility Theorem).\/} For the space $X = \{0, 1, 2\}$ take the following topology: $\mathcal{O} = \{ \{0,1,2\}, \{1\}, \{2\}, \{1,2\}, \emptyset\}$. Additionally, consider functions with codomain $\mathbb{R}$ or $\mathbb{S}^2$, equipped with the standard topology. For this topology, global continuous functions defined on $X$ will be constant. Let ${\bf Z} = \{z_1, z_2\}$ and let $\delta_i$ be the constant probability in $P(X)$ that assigns probability one to $z_i$ for $\{i = 1,2\}$. Preferences are defined as follows:
\[ \{1\} \Vdash \delta_2 \prec \delta_1 \]
\[ \{2\} \Vdash \delta_1 \prec \delta_2 \]
\[ \{1, 2\} \Vdash (\delta_1 \prec \delta_2) \vee (\delta_2 \prec \delta_1) \]
At point $0$, where $\{0, 1, 2\}$ is confirmed, $\delta_1$ and $\delta_2$ are not comparable. Preferences can be extended to $P(X)$. The typical element of $P(X)$ can be written
as ${p} = p_1\delta_1 + (1-p_1)\delta_2$, where $p_1$ is a continuous function over $X$ with values in $[0,1]$ (hence, given the topology, a constant function). Now, for
all ${p}$ and ${q}$,
\[ \{1\} \Vdash {q} \prec {p} \ \mbox{if and only if} \ q_1<p_1 \]
\[ \{2\} \Vdash {q} \prec {p} \ \mbox{if and only if} \ 1-q_1<1-p_1. \]
Clearly, for all $U$, $U \Vdash {q} \sim {p}$ when $p_1=q_1$, but when $p_1 \neq q_1$ $p$ and $q$ will not be comparable over all of $\{0,1,2\}$. As a consequence, it will now be shown,
there will not exist a global function that represents preferences. Each of the three assumptions can be confirmed to be satisfied for all open $U$. First consider $U=X$. There are no ${p}$ and ${q}$ with $X \Vdash {q} \prec {p}$ and so the asymmetry, negative transitivity, monotonicity, local character, independence, and continuity conditions are trivially satisfied (the antecedent conditions are never met). Now consider $\{1\}$. In this case, the conditions only need to be confirmed for standard real numbers $p_1$ and $q_1$ in $\mathbb{R}$ ordered by $<$. Asymmetry and negative transitivity are immediate. Monotonicity and local character are trivially satisfied, because the only subset of $\{1\}$ is $\emptyset$. Preferences satisfy independence because real numbers satisfy independence (if $q<p$ then for any $r \in [0,1]$ and $a \in (0,1]$ it must be that $aq + (1-a)r < aq + (1-a)r$). Along the same lines, it is easy to confirm that continuity is satisfied. The argument for $\{2\}$ is identical. It now remains to confirm that there can be no global continuous (hence, constant) functions that represent preferences. It is sufficient to  consider $\delta_1$ and $\delta_2$. There are no constant functions $f: X \rightarrow \mathbb{R}$ and $g: X \rightarrow \mathbb{R}$, which are not comparable at 0, have $f(1)<g(1)$, and $g(2)<f(2)$. $\blacksquare$

\bigskip

The difficulty relates to stages (like point 0 in the example above) where no two probability distributions are comparable. In the classical setting, if no two probabilities can be ranked by $\prec$ it follows that the decision-maker is indifferent between every pair of probabilities (by an application of the law of the excluded middle). This is not necessarily the case in the constructive setting. In this example,
preferences are not representable over $P(X)$. They are however representable over $P(\{1,2\})$ because $\{\{1,2\}, \{1,2\}, \{1,2\}, \emptyset\}$ is the discrete topology. Over $\{1,2\}$ it {\em is\/} possible to find continuous real-valued functions to represent preferences because all functions over $\{1,2\}$ are continuous. However, there is no way to extend these to continuous functions over $X$. One way to
interpret the result is that if preferences depend upon how a mathematical conjecture (e.g., the P vs. NP conjecture) is resolved then
it is not possible to commit to an assertion so long as the conjecture remains unresolved.\footnote{The example raises the question whether
the stages of knowledge in this example can be treated in the manner of states in Subjective Expected Utility theory, a la Savage \citet{savage1954}. This complex question and its answer are taken up in a separate paper.}

A second example is of the set $X=[0,2]$ with the standard topology and illustrates a different kind of failure. This relates to the part of the theorem asserting that the function which represents preferences is unique up to positive linear transformations.

\bigskip

\noindent {\bf Example 3 (Uniqueness Failure).\/}
In this example, let the support be ${\bf Z} = \{z_1, z_2\}$, and $X = [0,2]$ with the standard topology. Suppose $[0,1) \Vdash \delta_1 \prec \delta_2$ and
$(1,2] \Vdash \delta_2 \prec \delta_1$. Over any $U$ which contains $1$, $\delta_1$ and $\delta_2$ are not comparable (in the classical framework, $\delta_1 \sim \delta_2$ would
be deduced). Preferences can be extended to $P(X)$ in the obvious way, so that the distribution that gives the better prize with higher probability is preferred. Now when the standard construction from the expected utility theorem, using two fixed lotteries for calibration, is carried out, it creates a discontinuity at 1. This construction assigns utility 1 to the best option, utility 0 to the worst option, and for any other probability distribution sets utility equal to the probability of the better prize. Around $x=1$ all rankings flip. However, the failure of that particular construction does not preclude the possibility that there is some other construction of the representing function. For instance, in this example, it would appear that preferences can be represented by choosing continuous functions such as  $w(z_1) = x$ and $w(z_2) = 1$ for $x\in X$ (and the corresponding expected utility for all distributions in $P(X)$). In this case, the theorem fails for a different reason --- uniqueness up to positive linear transformations does not hold. To see this, consider another pair of functions: $v(z_1) = 0$ and $v(z_2) = \mbox{sgn}(1-x) \times (1-x)^2$ where sgn is the signum (sign) function. The function $v$ represents preferences as well, and so $w(\cdot)$ has to be a positive linear transformation of $v(\cdot)$. In other words,
there are continuous real-valued functions $a: X\rightarrow \mathbb{R}_+$ and $b: X\rightarrow \mathbb{R}$ with $w(\cdot) = a v(\cdot) + b$. In this case, it is possible to show that
$w(\cdot) = \left|\frac{1}{1-x} \right| v(\cdot) +x$ for all $x \neq 1$. However, since $a = \left|\frac{1}{1-x} \right|$, it diverges at $x=1$. So, while $v$ also represents preferences, there are no continuous real functions $a:[0,2]\rightarrow \mathbb{R}_+$ and $b:[0,2]\rightarrow \mathbb{R}$ such that $w(\cdot) = av(\cdot)+b$. $\blacksquare$

\subsection{The Constructive Expected Utility Theorem}

Two different approaches are taken to remedy the failure of the theorem. The first is to restrict preferences in a way that rules out the problematic situations. The second is to limit attention to sheaves where the law of the excluded middle holds. The next axiom, Assumption~4, provides the first remedy. It restricts preferences by limiting attention to $X$ where either $[(\forall p, q), p \sim q]$ or $[(\exists p, q), p \prec q]$ can be locally confirmed. In other words, the decision maker is either able to assert indifference between all distributions, or is able to assert strict preference between {\it at least\/} two choices. Note that these two conditions are negations of one another and it is required that one of the two hold. The axiom below is just an interpretation of this requirement using sheaf semantics \citep[][p.~316]{maclane1996}. The two conditions may not exhaust all possibilities, with the excluded situation being that of non-comparability: preferences are restricted to exclude the possibility that no two probability distributions are comparable. The axiom will be called {\em minimal comparability\/} -- it is satisfied if {\em just two\/} variable lotteries are comparable (the other lotteries are not required to be comparable).

\begin{axiom}[Minimal Comparability]
There exists an open covering $\{W_i\}$ of $X$ (with $\bigcup_i W_i=X$) such that, for each $W_i$, one of the following two conditions hold:
\begin{itemize}
\item For all $W_i^\prime \subseteq W_i$ and all $p_i$ and $q_i$ in $P(W_i^\prime), p_i \sim q_i.$
\item There are $ p_i, q_i \in P(W_i)$ with $ p_i \prec q_i.$
\end{itemize}
\end{axiom}

\begin{figure}
\begin{center}
\begin{tikzpicture}
  \draw[->] (0, 0) -- (4.5, 0) node[right] {$x$};
  \draw[->] (0, 0) -- (0, 4.5) node[above] {$y$};
  \draw[line width=0.2mm][scale=1, domain=0:4, smooth, variable=\x, black] plot ({\x}, {\x});
  \draw[line width=0.2mm][scale=1, domain=0:4, smooth, variable=\x, black] plot ({\x}, {4-\x});
  \draw[line width=0.2mm][scale=1, domain=0:4, smooth, variable=\x, black] plot ({\x}, {1+0.5*\x});
  \draw[line width=0.2mm][scale=1, domain=0:4, smooth, variable=\x, black] plot ({\x}, {3-0.5*\x});
  \draw[line width=0.2mm][scale=1, domain=0:4, smooth, variable=\y, black]  plot ({\y}, {2});
  \draw [dashed] (2,0) -- (2,2)node[color=black] at (2, -0.25) {$\hat{x}$};
  \node [color=black] at (4.2,4.2) {$\delta_1$};
  \node [color=black] at (4.2,0.2) {$\delta_5$};
  \node [color=black] at (4.2,2.2) {$\delta_3$};
  \node [color=black] at (4.2, 3.2) {$\delta_2$};
   \node [color=black] at (4.2,1.2) {$\delta_4$};
\end{tikzpicture}
\caption{Failure of minimal comparability}
\end{center}
\end{figure}

In classical proofs of the theorem the {\it law of the excluded middle\/} ensures that this claim is always true.
The assumption holds if one or the other condition holds globally, but could also hold with one condition holding in some region of $X$,
while the other condition holds in another disconnected region. If $[(\forall p, q), p \sim q]$ over $W_1$ and $[(\exists p, q), p \prec q ]$ over $W_2$ then $W_1$ and $W_2$ could not have a non-empty intersection (because both conditions cannot hold simultaneously).
The excluded situation is illustrated in Figure~3 for the case of real functions ordered by $<$. The lines are the valuation of five prizes as a function of $x$. Compare $\delta_1$ and $\delta_5$.
There is no open containing $\hat{x}$, that does not also contain open subsets with $\delta_1<\delta_5$ in one subset, and $\delta_5<\delta_1$ in another. Hence, $\delta_1$ and $\delta_5$ are not comparable at $\hat{x}$. The same is true for any two $\delta_i$ and $\delta_j$, and so any $p, q \in P(W)$. In the interpretation of open sets as observations, at $\hat{x}$, any observation leaves open the possibility that more refined observation may find $p\prec q$ or $q\prec p$, and this is true for every pair of distributions.

The second approach to obtaining a representation result will be by assuming that the sheaf satisfies excluded middle ($\forall \phi, \phi \vee \neg \phi$), making the logic classical.
In the present setting this translates to:
\begin{axiom}[Classical Logic]
For every open set $W$ in $\mathcal{O}(X)$, $W \cup \mbox{Int}(W^c) = X$, where $\mbox{Int}(W^c)$ denotes the interior of $W^c$. Equivalently, if $W$ is open, $W^c$ is
also open.
\end{axiom}
The truth value of any proposition $\phi$ is identified with an open set $W$ in $\mathcal{O}(X)$. The negation of $\phi$, i.e., $\neg \phi$, has as its truth value Int($W^c$),
which is the largest open set in $X$ that is disjoint from $W$. This axiom asserts that at every point in $X$ either $\phi$ or $\neg \phi$ holds. An example of a topology where
Assumption~5 holds is the discrete topology (e.g., consider $X=\{1,2\}$ and $\mathcal{O}(X) = \{ \{1, 2\}, \{1\}, \{2\}, \emptyset \}$).

Two versions of the expected utility theorem can now be proved. Recall that the classical theorem starts with $\prec$ defined on $\mathbb{S}^n$ and states that
(counterparts of) Assumptions~1-3 are satisfied if and only if
there is a real-valued affine function $u:\mathbb{S}^n \rightarrow \mathbb{R}$ that ``represents preferences" in the sense that
\[ p \prec q \Longleftrightarrow \sum_{z=1}^n u(\delta_z)p_z < \sum_{z=1}^n u(\delta_z)q_z.\]
A constructive version of this result is first proved using Assumption~4. Here, Assumptions~1-4 provide sufficient conditions for there to exist continuous functions that represent preferences. Then, a version of the result using Assumptions~1-3 together with Assumption~5 is proved. This theorem provides both necessary and sufficient conditions, but the result becomes non-constructive. However, with the classical logic assumption, expressive phenomena such as rankings that are non-comparable, or undetermined,  disappear.

The theorem below states that something like the classical theorem continues to be true. It is important to note that the theorem only provides a {\it local representation\/} in the sense
that there is a covering of $X$ by $\{W_i\}$ and a family of continuous functions $\{u_{W_i}\}$ where for each $f\in P(W_i), u_{W_i}(f)$ is a {\it partial\/} continuous function representing preferences over $W_i$. Alternatively stated, at each $x\in X$, there is an open neighborhood $W_x$ of $x$ where preferences for $f$ can be represented by a continuous real-valued function $u_{W_x}(f)$. This still leaves open the question whether there is a single continuous function $u(f)$ over all of $X$ such that each $\{u_{W_i}(f)\}$ is just the restriction of $u(f)$ to $W_i$. This question will be addressed below in Corollary~1.

\newtheorem{thm}{Theorem}
\begin{thm}[Constructive Expected Utility Theorem]
Suppose $X$ is a topological space, with $\mathcal{O}(X)$ its collection of open sets and $P$ a sheaf of variable lotteries. A binary relation $\prec$ is defined on $P$ and satisfies Assumptions 1--4. Then there is an open covering $\{W_i\}$ of $X$ (with $\bigcup_i W_i=X$), and there exist mappings $u_{W_i}: P(W_i) \rightarrow {\bf R}(W_i)$ such that for $p, q\in P(W_i)$
\begin{equation}
W_i \Vdash p \prec q \quad \Longleftrightarrow \quad \sum_{z\in \Phi(W_i)} u_{W_i}(\delta_z|_{W_i})p_z < \sum_{z\in \Phi(W_i)}^n u_{W_i}(\delta_z|_{W_i})q_z,
\end{equation}
where $\Phi(W_i) \subseteq \{1, 2, \ldots n\}$.
Moreover, the $u_{W_i}$ are unique up to positive linear transformations. Additionally, suppose $u_W$ exists and represents preferences over $W$ and, for $V \subseteq W$, $u_V$ represents preferences over $V$. If $\rho_{WV}$ is the restriction from $P(W)$ to $P(V)$ and $\rho_{WV}^\prime$ is the restriction from ${\bf R}(W)$ to ${\bf R}(V)$, then
$ u_V \circ \rho_{WV} = \rho_{WV}^\prime \circ u_W $.
\end{thm}

From the definition of indifference above, it follows that,
\begin{equation}
W_i \Vdash p \sim q \quad \Longleftrightarrow \quad \ \sum_{z\in \Phi(W_i)} u_{W_i}(\delta_z|_{W_i})p_z = \sum_{z\in \Phi(W_i)}^n u_{W_i}(\delta_z|_{W_i})q_z.
\end{equation}
This is because strict inequality ($<$) and equality ($=$) for real continuous functions in the sheaf {\bf R}($X$) is defined analogously to $\prec$ and $\sim$.
So, $\neg( p \prec q)$ and $\neg( q \prec q)$ imply $\neg(\sum_{z\in \Phi(W_i)} u_{W_i}(\delta_z|_{W_i})p_z < \sum_{z\in \Phi(W_i)} u_{W_i}(\delta_z|_{W_i})q_z)$ and
$\neg(\sum_{z\in \Phi(W_i)} u_{W_i}(\delta_z|_{W_i})q_z < \sum_{z\in \Phi(W_i)} u_{W_i}(\delta_z|_{W_i})p_z)$. Equation~2 follows.

\begin{figure}
\begin{center}
\begin{tikzcd}
W                & P(W) \arrow[dd, "\rho_{WV}"'] \arrow[rr, "u_W"] &  & {\bf R}(W) \arrow[dd, "\rho^\prime_{WV}"] \\
                   &                                                 &  &                                     \\
V \arrow[uu, hook] & P(V) \arrow[rr, "u_V"]                          &  & \bf{R}(V)
\end{tikzcd}
\caption{Commutative diagram summarizing the results in Theorem~1 and Corollary~1 when $P(W)$ is equipped is
with a binary relation $\prec$ satisfying Assumptions 1-4.}
\end{center}
\end{figure}

\begin{figure}
\begin{center}
\begin{tikzcd}
                                                                    &  &                                                                   & {\bf R}(W) \arrow[dd, "\rho_{WV}^\prime"] \\
P(W) \arrow[dd, "\rho_{WV}"'] \arrow[rr, "u_W"] \arrow[rrru, "v_W"] &  & {\bf R}(W) \arrow[dd, "\rho^\prime_{WV}"] \arrow[ru, "\phi_W!"] &                                     \\
                                                                    &  &                                                                   & \bf{R}(V)                                \\
P(V) \arrow[rr, "u_V"] \arrow[rrru, "v_V", shift left]              &  & {\bf R}(V) \arrow[ru, "\phi_V!"]                                &
\end{tikzcd}
\caption{The uniqueness result. $P(W)$ is equipped is with a binary relation $\prec$ satisfying Assumptions 1-4.}
\end{center}
\end{figure}

Several observations are in order. First, note that $u_{W_i}$ is a mapping from the space of variable lotteries $P(W_i)$ to the space of continuous real-valued
functions {\bf R}($W_i$). The real-valued utility of a lottery is defined pointwise over $W_i$. This may vary even for a constant lottery $\delta_z$ because
preferences are allowed to vary. The situation is illustrated in the commutative diagram in Figure~4. Note, in particular, that $u_W$ is defined for any $W\in \mathcal{O}(X)$ for variable lotteries in $P(W)$,
and for any $V \subseteq W$ the restriction to $P(V)$ followed by the representation of preferences in ${\bf R}(V)$
yields the same result as first representing preferences in ${\bf R}(W)$, and then taking the restriction to ${\bf R}(V)$.
I.e., $ u_V \circ \rho_{WV} = \rho_{WV}^\prime \circ u_W $. The uniqueness result is depicted by the commutative diagram in Figure~5.
If there is any other collection of functions $v_W$ that represent preferences, there are unique functions (positive linear transformations)
$\phi_W:{\bf R}(W) \rightarrow {\bf R}(W)$ and $\phi_V:{\bf R}(V) \rightarrow {\bf R}(V)$ that make the diagram commute.
A second observation is that the $W_i$ are chosen in such a way that there exist some pair of distinct and comparable distributions (from
Assumption~4). However, there may be many other distributions that are not comparable over $W_i$. Further refinement of $W_i$, by taking a
suitable covering, may reveal new rankings. For instance, a pair $r$ and $s$ may not be comparable over $W_i$, but may be comparable over some $V\subset W_i$. If
$r|_V \prec s|_V$, it follows that $\sum_{z\in \Phi(W_i)} u_{W_i}(\delta_z)|_{W_i}|_{V}r_z|_{V} < \sum_{z=1}^n u_{W_i}(\delta_z)|_{V}s_z|_{V}$ holds over $V$.
The decision-maker at stage $x\in X$ can assert all the open sets $V$ that contain $x$, and if there is some $V$ over which $r\prec s$, then
the corresponding expected utility comparison of $r$ and $s$ can be made over $V$ as well. Third, in addition to the representation being order-preserving,
non-comparability is preserved as well. If, over any set $V \subseteq W_i$, $r$ and $s$ are not comparable, then it is also the case that $u_{W_i}(r)|_V$ and $u_{W_i}(s)|_V$ are not comparable over $V$. Finally, the previous theorem only provides a local representation by partial functions. The reason can be understood from an examination of the condition in Assumption~4: $W \Vdash (\exists p, q), p \prec q$ requires that there be a covering $\{W_i\}$ of $W$ such that over each $W_i$ there are at least two distributions between which the decision-maker can express strict preference. But it is possible to have the following situation:
$W = W_1 \cup W_2$ together with $W_1 \Vdash p\prec q$ and $W_2 \Vdash r \prec s$ for $r$ and $s$ different from $p$ and $q$.
Further, it could be that $p$ and $q$ are not comparable over $W_2$ and $r$ and $s$ are not
comparable over $W_1$. The different $u_{W_i}$ may not then constitute a compatible (matching) family collating to a single function. The following assumption ensures that the local representing functions $u_{W_i}(p|_{W_i})$ can be collated into a single function.

\begin{axiom}
There exists a covering $\{W_i\}$ of $X$, and two compatible families $\{p_i:W_i \rightarrow \mathbb{S}^n\}$ and $\{q_i:W_i \rightarrow \mathbb{S}^n\}$ collating to $p:X \rightarrow \mathbb{S}^n$ and $q:X \rightarrow \mathbb{S}^n$  respectively, satisfying one of the following two conditions for each index $i$:
\begin{itemize}
\item $W_i \Vdash p_i \prec q_i$
\item For all $W_i^\prime \subseteq W_i$ and all $p_i$ and $q_i$ in $P(W_i^\prime), p_i \sim q_i.$
\end{itemize}
\end{axiom}

This yields the following corollary of Theorem~1.

\newtheorem{cor}{Corollary}
\begin{cor}
Suppose Assumptions 1-3, and Assumption 6 hold. Then the following holds for all open $W \subseteq X$.
\begin{enumerate}
\item For all $p, q \in P(W)$ there is an open covering $\{W_i\}$ of $W$ and affine
mappings $u_{W_i}: P(W_i) \rightarrow {\bf R}(W_i)$ such that Equation~1 holds. Further, the functions $\{ u_{W_i}(p|_{W_i}) \}$ constitute a compatible family in $\{{\bf R}(W_i)\}$ that can be collated to
obtain an element $u_W(p)$ of ${\bf R}(W)$. For any $V\in \mathcal{O}(W)$, $u_{V}(p|_V)$ can be obtained as the restriction of $u_W(p)$ to $V$.  As before, $u_W$ is unique up to positive linear transformations.
\item Suppose for $W \subseteq X$, $u_W$ represents preferences over $W$ and, for $V \subseteq W$, $u_V$ represents preferences
over $V$. Let $\rho_{WV}$ be the restriction from $P(W)$ to $P(V)$ and
$\rho_{WV}^\prime$ the restriction from ${\bf R}(W)$ to ${\bf R}(V)$. Then $u_V  \circ \rho_{WV}  = \rho_{WV}^\prime \circ u_W$.
\end{enumerate}
\end{cor}
When $p\in P(X)$, $u_X(p) \equiv u(p)$ is a global element of ${\bf R}(X)$.
The important difference with the previous theorem is that the $u_W$ now satisfy the gluing condition for sheaves.
The mappings $u_W$ defined from $P(W)$ to ${\bf R}(W)$ for all $W \in \mathcal{O}(X)$ and satisfying the gluing condition constitute what is known as a{\em natural transformation\/} of sheaves.

It will be useful to revisit Examples~2 and 3 in light of the results of this section. In the case of Example~3, the difficulties are resolved by the addition of two distinct variable lotteries
(as in Assumption 6). The two lotteries can be used for calibration of the other lotteries.\footnote{It may be the case that there are lotteries that are sometimes not comparable with one or the
other of these lotteries, but in that case the two will yield other lotteries, as their weighted averages, that can be used for calibration. This point will be evident from an inspection of the proof.} The end result is that there is a global mapping $u(p)$, unique up to positive linear transformations, that restricts to any domain $W$ of a variable lottery $p$ and represents preferences
over $W$. Of course, for $W \subset X$ there may be new lotteries in $u_W$ in addition to the restriction $p|_W$ of lotteries $p\in P(X)$.
Example~2 is more challenging. In this case, $X = \{0, 1, 2\}$ with open sets $\{\{0, 1, 2\}, \{1\}, \{2\}, \{1, 2\}, \emptyset \}$. The preferences defined have $\{1\} \Vdash \delta_1 \prec \delta_2$ and $\{2\} \Vdash \delta_2 \prec \delta_1$. A mapping $f$ that represents preferences must have $f(\delta_1)<f(\delta_2)$ at 1 and $f(\delta_2)<f(\delta_1)$ at 2. Restricted to $\{1, 2\}$ any function is continuous, but there is no way of extending $f$ to 0 for $\delta_1$ and $\delta_2$. This is because there can be no continuous function $f$ that represents preferences since, with this topology, continuous real-valued functions must be constant. So, $f(\delta_1)$ and $f(\delta_2)$ must be undefined at 0. However, this does not contradict Theorem~1 or Corollary~1 above because, after adding $\delta_3$ and $\delta_4$ with $\{0,1,2\} \Vdash \delta_3\prec \delta_4$, preferences cannot satisfy the axioms if $\delta_1$ and $\delta_2$
can be numerically represented at $0$. Specifically, the negative transitivity condition fails now. To see this, note that any two rational functions $v, w \in {\bf Q}(\{0, 1, 2\})$ must be constant. Denote their value at any point in $\{0, 1, 2\}$ by $\hat{v}$ and $\hat{w}$ respectively. For any $f$ that represents preferences it must be that $f(\delta_1)<f(\delta_2)$ at 1 and $f(\delta_2)<f(\delta_1)$ at 2. Since $\{0, 1, 2\} \Vdash \delta_3\prec \delta_4$, then from negative transitivity it must follow that $\{0, 1, 2\} \Vdash \delta_1 \prec \delta_4$ or $\{0, 1, 2\} \Vdash \delta_3 \prec \delta_1$. Assume the former (the second case being similar). The mapping $f$ can be
chosen such that $f(\delta_4)=1$ and $f(\delta_1)=0$, so that $f(\delta_2)(2)<0$ and $f(\delta_2)(1)>0$. Now choose $0 < \hat{v} < \hat{w} <  f(\delta_2)(1)$. The lotteries
$ v\delta_4+(1-v)\delta_1$ and $w\delta_4+(1-w)\delta_1$ have $ v\delta_4+(1-v)\delta_1 \prec w\delta_4+(1-w)\delta_1$ (from monotonicity, which is implied by independence) but neither is comparable to $\delta_2$, violating negative transitivity.\footnote{A closely related argument applies in the general case as well relying, as here, on the fact that the upper and lower Dedekind cuts of any real-valued
function are arbitrarily close (see the proof of Lemma~2).}

\subsection{The Classical Expected Utility Theorem}
An alternative to employing Assumption~4 to obtain the representation is to restrict attention to topologies on $X$ that are ``classical" and satisfy the law of the excluded middle. This is done by requiring that $\mathcal{O}(X)$ satisfy Assumption~5, yielding the following theorem. This theorem can be stated for $P(X)=S(X)$ (in other words, for $\Phi(W) = \{1, 2, \ldots n\}$ for all $W\subseteq X$), because
the complication of partially defined preferences does not arise.

\begin{thm}[Expected Utility Theorem with Classical Logic]
Suppose $X$ is a topological space, with $\mathcal{O}(X)$ its collection of open sets that satisfies Assumption~5. $P$ is a sheaf of variable lotteries with a binary relation $\prec$ is defined on it. The binary relation $\prec$ satisfies Assumptions 1--3 if and only if there is an open covering $\{W_i\}$ of $X$ (with $\bigcup_i W_i=X$), and there exists a mapping $u: P(X) \rightarrow {\bf R}(X)$ such that,
\begin{equation}
W_i \Vdash p|_{W_i} \prec q|_{W_i} \quad \Longleftrightarrow \quad \sum_{z=1}^n u(\delta_z)|_{W_i}p_z|_{W_i} < \sum_{z=1}^n u(\delta_z)|_{W_i}q_z|_{W_i}.
\end{equation}
Moreover, $u$ is unique up to positive linear transformations. Additionally, suppose $u|_W$ represents preferences over $W$ and, for $V \subseteq W$, $u|_V$ represents
preferences over $V$. If $\rho_{WV}$ is the restriction of $P(W)$ to $P(V)$ and $\rho_{WV}^\prime$ is the restriction of ${\bf R}(W)$ to ${\bf R}(V)$, then $u|_V \circ \rho_{WV}   = \rho_{WV}^\prime \circ u|_W $.
\end{thm}

Since the two conditions in Assumption~4 are negations of one another, Assumption~5 implies Assumption~4. Consequently, the sufficiency part of the proof is immediate. Despite its superficial similarity, Theorem~2 is qualitatively very different from Theorem~1. Theorem~1 and its corollary require at least two distributions to be comparable (by $\prec$ or $\sim$), and Assumption~4 provides such distributions. Assumption~6 makes {\it every\/} pair of distributions comparable. Let $ \llbracket \phi \rrbracket \in \mathcal{O}(X)$ denote the truth value of proposition $\phi$ (i.e., the stages where it holds). Then, for any $p$ and $q$, by the law of the excluded middle we have $ \llbracket p \prec q \rrbracket \cup \llbracket \neg (p \prec q) \rrbracket=X$. Non-comparability disappears as a phenomenon. The ability to describe incomplete stages of knowledge, where it is not possible to assert either one of (1)~a proposition $\phi$, or (2)~its negation, $\neg \phi$, is also lost. It is possible to make the logic two-valued. This requires choosing some $x\in X$ and replacing every lottery with the equivalence class of lotteries between which the decision-maker is indifferent at $x$. One can induce global preferences between these classes based on preference at $x$. Now $\{X, \emptyset\}$ are the only possible truth values.

In classical decision theory, there is a separation of the basis of preference into beliefs (probability) and desires (utility).
One might ask if there is a version of the theory where beliefs are variable, but preferences are not.
In this vein, it is possible to restrict preferences so that there is variation only in probabilities and not in utilities. One way of doing this is by requiring that comparisons of {\em constant\/} probabilities in $P(X)$ only take truth values $X$ or $\emptyset$.
The constant probabilities are of the form $p = \sum_z p_z\delta_z \in P(X)$ where
the $p_z:X\rightarrow [0,1]$ are constant functions. The following axiom requires that {\it every\/} pair of such probabilities is comparable over $X$ and one of $p\prec q$, $q\prec p$, or $p\sim q$ is either always true or never true.
The requirement is that if a decision-maker prefers $q$ to $p$ at some $x\in X$, he or she must do so at every $y\in X$.
Beliefs about the likelihood of the different elements of $Z$ are still allowed to change, but the valuation of the outcome in $Z$ does not change. As before, let $\llbracket \phi \rrbracket$ denote the set in $\mathcal{O}(X)$ where proposition $\phi$ can be asserted
to be true. The required assumption is:
\begin{axiom}[Comparison of constant probabilities]
For all constant distributions $p$ and $q$ in $P(X)$
\[ \llbracket p \prec q \rrbracket \in \{ X, \emptyset\}, \llbracket q \prec p \rrbracket \in \{ X, \emptyset\}, \llbracket p \sim q \rrbracket \in \{ X, \emptyset\}.\]
Additionally, $ \llbracket p \prec q \rrbracket \cup \llbracket q \prec p \rrbracket \cup \llbracket p \sim q \rrbracket = X$.
\end{axiom}

\begin{cor}
Suppose Assumptions~1-3 and Assumption~7 hold. Assume also that there are two distinct degenerate distributions (labeled $\delta_1$ and $\delta_n$) with $\delta_1 \prec \delta_n$ and, for all {\it i\/},  $\delta_1 \precsim \delta_i \precsim \delta_n$.
Then, there is a $u: P(X) \rightarrow {\bf R}(X)$ such that
\[ X \Vdash  p \prec q \quad \Longleftrightarrow \quad \sum_{z=1}^n u(\delta_z)p_z < \sum_{z=1}^n u(\delta_z)q_z, \]
holds for constant functions $u(\delta_z):X\rightarrow \mathbb{R}$.
\end{cor}
The assumption that $\delta_1 \prec \delta_n$ and $\delta_1 \precsim \delta_i \precsim \delta_n$ over $X$ is just a variation on
Assumption~6 for the current setting where all constant distributions are always globally comparable. Positive linear transformations could make
these functions non-constant (e.g. $v = au+b$ and $a$ or $b$ are non-constant functions of $x$), so uniqueness holds only for constant $a$ and $b$.

\subsection{Example 4}
Examples 2 and 3 illustrated a number of important features of the framework of variable lotteries. The example presented next shows how the framework
encapsulates another idea -- using solutions of simpler problems to patch together solutions of more complicated problems. This idea is already present in the classical theorem,
and the intent with this final example is to articulate the idea within the current framework and give it clearer expression. Many expositions, including \citet{pratt1964},
that present expected utility theory as a normative guide to behavior rely on the idea that the choice between multiple complicated lotteries can be simplified by the calibration
exercise present in the expected utility construction.

\begin{figure}[htbp]
\centering
  \begin{subfigure}{.48\textwidth}
   \centering
    \begin{adjustbox}{max width=\linewidth}
           \begin{tikzpicture}[scale=0.5]
  \coordinate (A) at (0,0);
  \coordinate (B) at (4,0);
  \coordinate (C) at (2,3);

  \fill[gray!30] (A) -- (B) -- (C) -- cycle;

  \draw (A) -- (B) -- (C) -- cycle;

  \node at (A) [below left] {$a$};
  \node at (B) [below right] {$b$};
  \node at (C) [above] {$c$};

  \foreach \point in {A, B, C}
    \fill [black] (\point) circle (2pt);
    \end{tikzpicture}
      \end{adjustbox}
    \caption{The simplex $\mathbb{S}^3$.}
    \label{fig:sub2}
  \end{subfigure}
  \begin{subfigure}{.48\textwidth}
   \centering
    \begin{adjustbox}{max width=\linewidth}
    \begin{tikzpicture}[scale=1.2]
\node (A) at (0,0) {\em a};
\node (B) at (4,0) {\em b};
\node (C) at (2,3.46) {\em c}; 

\node (AB) at ($(A)!0.5!(B)$) {\em ab};
\node (BC) at ($(B)!0.5!(C)$) {\em bc};
\node (CA) at ($(C)!0.5!(A)$) {\em ac};

\node (Center) at (barycentric cs:A=1,B=1,C=1) {\em abc};

\draw[->] (A) -- (AB);
\draw[->] (A) -- (CA);
\draw[->] (B) -- (BC);
\draw[->] (B) -- (AB);
\draw[->] (C) -- (CA);
\draw[->] (C) -- (BC);

\draw[->] (AB) -- (Center);
\draw[->] (BC) -- (Center);
\draw[->] (CA) -- (Center);

\end{tikzpicture}
    \end{adjustbox}
    \caption{The face incidence graph/poset.}
    \label{fig:sub1}
  \end{subfigure}
\caption{The simplex and corresponding face incidence graph/poset for Example 4.}
  \label{fig:test}
\end{figure}

For the purposes of this example, consider the space of lotteries $\mathbb{S}^n$ (probability distributions over $n$ prizes). This is a simplex and, like all simplices,
has the property that its boundaries are lower dimensional simplices. These constitute the {\em faces\/} of the simplex (i.e., simplices whose vertices are a subset of the vertices of the
original simplex). In Figure~6 the space of lotteries is depicted for $n=3$ on the left as a ``filled-in'' triangle, with each point denoting a probability distribution.
On the right is the corresponding face incidence graph/poset. This graph has as its nodes all possible simplices, with the full space of lotteries (the filled-in triangle $abc$) in the
center. The lower-dimensional simplices ($ab, ac, bc, a, b, c$) are also depicted. The arrows emanating from a simplex point to the simplices it is a face of. The intersection of any two
faces can be seen to be a (lower-dimensional) face of the simplices.  The structure in Figure~6 is a special case of a structure known as an {\em abstract simplicial complex\/} \citep[][p. 284]{rosiak2022sheaf}). \citet{rosiak2022sheaf} uses the notation $\sigma \rightsquigarrow \tau$ to indicate that $\sigma$ is attached to $\tau$ as a face. He goes on to define a {\em cellular
sheaf\/} $F$ of vector spaces on a face incidence graph $X$ which is (a)~an assignment of a vector space $F(\sigma)$ to each simplex $\sigma\in X$, together with (b)~a linear
transformation $F_{\sigma \rightsquigarrow \tau}: F(\sigma) \rightarrow F(\tau)$ for each arrow $\sigma \rightsquigarrow \tau$ satisfying $F_{\sigma \rightsquigarrow \sigma} = id$,
and if $\rho \rightsquigarrow \sigma \rightsquigarrow \tau$ then $F_{\rho \rightsquigarrow \tau} = F_{\sigma \rightsquigarrow \tau} \circ  F_{\rho \rightsquigarrow \sigma}$.\footnote{As
it turns out, it is unnecessary to spell out the data of the topology, covers, and sheaf conditions. The above definition is sufficient and contains all the data of a sheaf \citep[][p. 287]{rosiak2022sheaf}. Topologies can however be defined as the collection of principal upper (or lower) sets for the face incidence relation between simplices in the face incidence
graph.}

For the present context, it is possible to define a cellular sheaf $G$ where, for $\sigma\in X$ , $G(\sigma)$ is an ordered space of probability distributions satisfying the weak order and other axioms, and the restriction map $G_{\sigma \rightsquigarrow \tau}$ from $G(\sigma)$ to $G(\tau)$ is order-preserving. So, $G(a)$ would be a degenerate distribution, $G(ab)$ would equal $\{p\delta_a + (1-p)\delta_b| p\in [0,1]\}$, and so on. Rather than focusing on the construction of the expected utility representation, the properties of
the representation using a cellular sheaf are now described. The
definition of a cellular sheaf of vector spaces above is modified by limiting $F(\sigma)$ to be an interval $[\underline{u},\overline{u}]\subset \mathbb{R}$, where possibly $\underline{u}=\overline{u}$ (in either case this is a simplex, but of different dimensions). For $x_\sigma\in \sigma$, $F(x_\sigma) \in F(\sigma)$ is an indicator of preference for the lottery in $G(\sigma)$ corresponding to $x_\sigma$. The restriction mappings $F_{\sigma \rightsquigarrow \tau}$ will be positive
linear transformations. When $\sigma \rightsquigarrow \tau$ these transformations ensure that all the faces that attach to $G(\tau)$ have their expected utilities transformed such that they
have a common zero and unit in the simplex $F(\tau)$. Thus, a numerical indicator of preference for a lottery in a simplex $G(\sigma)$ is mapped to a numerical indicator for the same lottery in the simplex $G(\tau)$. For the example in Figure~6, $F(ab)$, $F(bc)$ and $F(ac)$ cannot all have the same lotteries that serve as the zero and unit, but in $F(abc)$ the zero and unit of one of the three can be adopted as the common zero and unit. Positive linear transformations now provide the harmonization, after which the expected utility function can be extended to
all of $G(abc)$.

\section{Conclusion}
This paper develops a model of preference and choice over a sheaf of variable lotteries. These lotteries vary over a topological space, which is chosen to encode
observability or verifiability constraints. Several results are proved. First, the classical expected utility theorem is shown to be non-constructive. Second, a constructive version
of the theorem for variable lotteries is proved, incorporating a distinction between indifference and non-comparability. Preferences defined on variable lotteries can be represented via a natural transformation that maps variable lotteries to the sheaf of continuous real-valued functions. In the earlier analogy, sheaves were motivated as probes from test spaces to some unknown space.
The natural transformation has the effect of turning probes of the space of lotteries into probes of the real numbers. Continuous real-valued functions are indicators of preference for variable lotteries and these functions faithfully reflect the ``inability to compare'' phenomenon. The theory includes a notion of refinement of information, whereby new rankings emerge from more detailed information. The strategy for the proof involves lifting the construction of the classical expected utility to the category of sheaves. Remarkably, suitably modified versions of the same classical axioms suffice for the result. Third, a version of the classical theorem is obtained by imposing a condition on the collection of open sets of the topology which has the effect of making the logic classical. Fourth, conditions are provided whereby local representations of preference over open sets can be collated to provide a global representation. Different observations, measurements, or computations might lead to different assertions about preferences, and the theory provides a way to combine such ``local'' assertions about preference into ``global'' assertions. The use of the word constructive in the title of the paper is used to suggest such a progression and ``building up'' of preferences from its constituent parts.

It is worth noting that the main theorem is extremely general. It is a generalization of the classical theorem in the sense that an additional condition on the topology (or the choice of the singleton space) leads to the classical result. It was suggested earlier that topologies can be viewed as observation systems. This was illustrated with concrete cases in Examples 2, 3 and 4. Several other choices are possible, and this is an avenue for further exploration. \citet{vickers1996} discusses other potential topologies, such as the Scott topology, which is suitable when Example~2 is extended to infinitely many stages of knowledge. Example~4 touches upon the idea of piecing together preferences in simple choice situations into preferences between complex alternatives. It should be possible to use the tools of this paper to treat this question in a more general way by considering more general ``test spaces'' and ``probes''.

The stages or points over which variable lotteries are defined raise the question of treating them like the states of subjective expected utility theory in the manner of Savage \citep{savage1954}.
This is not entirely straightforward, but some results are possible for the {\it subjective\/} expected utility theorem with lotteries defined over states. Indeed, the original motivation for the present research program was to define subjective probabilities over states of deductive knowledge. This problem and its solution will be taken up in a separate paper.

The constructive proof of the expected utility theorem in Sh($X$) suggests that it should be possible to carry the theorem over to other mathematical universes.
This is because of a strong robustness property of constructive theorems. Just as the constructive expected utility theorem was lifted to the category (topos) Sh($X$), it can be
carried over to other mathematical universes that include the real number object. One such is the topos of smooth worlds \citep[see][]{bell2008} in which it is possible to introduce
infinitesimals into the preference and choice setting in a meaningful way. There may be other possibilities as well.

\printbibliography

\newpage
\section*{Proofs}
A series of claims is first proved. They will be familiar from their counterparts in the classical setting. In some instances the proofs are direct adaptations of textbook versions
of the Expected Utility Theorem (e.g. \citep{kreps1988} and \citep{fishburn1970}). However, sometimes they are very different (e.g., for solvability, in Lemma~2).
\newtheorem{claim}{Lemma}
\begin{claim}[Monotonicity]
Suppose $p,q \in P(W)$, and $a:W\rightarrow [0,1]$ and $b:W\rightarrow [0,1]$ are continuous functions with $W \Vdash a<b$. Then,
\begin{equation}
W \Vdash p \prec q \Rightarrow aq+(1-a)p \prec bq + (1-b)p.
\end{equation}
\end{claim}
\noindent {\it Proof.\/}
This follows from the fact that for each $x\in W$, $a(x)q(x)+(1-a(x))p(x)$ is a convex combination of $b(x)q(x) + (1-b(x))p(x)$ and $p(x)$. In particular, over $W$,
$aq+(1-a)p = t(bq + (1-b)p) + (1-t)p$ for $t=a/b$, a continuous function that maps $W$ to $[0,1)$ . By Assumption 2, $p \prec bq+(1-b)p$ (since $p = bp + (1-b)p$).
So, by another application of Assumption 2,  $t(bq+({1}-b)p) +({1}-t)p \prec bq+({1}-b)p$. The term on the left is just $aq+({1}-a)p$, which yields the conclusion in equation~(4).
This proves the claim. $\blacksquare$

The next lemma is a version of the classical result which is sometimes called solvability. The classical results asserts that for any three lotteries $p, q,$ and $r$ with $r \precsim q \precsim p$
and $r\prec p$, there exists a real number $a^*$ such that $q\sim a^*p+(1-a^*)r.$ The result below is for variable lotteries and states the result for a continuous function $a^*$.
A notewothy feature of the proof is the reliance on the Dedekind construction of the real numbers to arrive at $a^*$. This, in turn, relies on the fact that in the category of sheaves, Sh($X$), the Dedekind reals are isomorphic to the set of continuous real-valued functions. Parts of the proof are patterned after \citet[][Section VI, Theorem~2]{maclane1996}, which provides a proof of this isomorphism.\footnote{See Appendix 1. Unlike the category {\bf Set\/}, the Dedekind reals are distinct from the Cauchy reals in Sh($X$).}

\begin{claim}[Solvability]
Suppose $p,q,r \in P(W)$, with $ W \Vdash r \precsim q \precsim p$ and $r\prec p$. Then there exists a unique continuous function $a^*: W \rightarrow [0,1]$ such that $$W \Vdash q\sim a^*p+({1}-a^*)r.$$
\end{claim}
\noindent {\it Proof.\/}
The function $a^*$ will be defined in terms of two sets of functions, $L(W)$ and $U(W)$. These are defined below in terms of $\prec$ and correspond to lower and upper Dedekind cuts for the present context of a continuous function. Recall that {\bf Q}($W$) was defined to be the set of locally constant {\em rational-vaued\/} functions with domain $W$.\footnote{In contrast to continuous real-valued functions, functions in {\bf Q}($W$) are always comparable: for $v$ and $w$ in {\bf Q}($W$)
it is always the case that one of $v<w$, $w<v$ or $w=v$ holds at every $x\in W$ (i.e., holds everywhere in a neighborhood of $x$).} $L(W)$ and $U(W)$ are subsets of {\bf Q}($W$) and
defined as follows:
\begin{equation}
\begin{aligned}
 L(W) = \{v \in {\bf Q\/}(W)| & (\forall x\in W)[ (v(x)\in [0,1]\  \mbox{and}\ \\
                                                &  {v(x)}p(x)+({1}-{v(x)})r(x) \prec q(x))\  \mbox{or}  \  v(x)<{0}] \}
   \end{aligned}
 \end{equation}

\begin{equation}
\begin{aligned}
U(W) = \{v \in {\bf Q\/}(W)| & (\forall x\in W) [(v(x)\in [0,1]\  \mbox{and} \  \\
                                               & q(x) \prec {v(x)}p(x)+({1}-{v(x)})r(x))\  \mbox{or} \  v(x)>{1}] \}
\end{aligned}
\end{equation}

Note that $L$ and $U$ are constructed as the union of two components. For $L$, the first component is the set of rationals $v$ over $W$ for which $vp+({1}-{v})r \prec q$, while the second component is the set of negative rational numbers. For $U$, the first component is the set of rationals $v$ over $W$ for which $q\prec ap+({1}-{v})r$ and the second component is the set of rationals greater than one. In each of 1--5 below, a Dedekind condition is stated first, followed by the corresponding condition that needs to be proved in the present context.

\begin{description}[style=unboxed, leftmargin=0cm]
  \item[1. Inhabited.] $W \Vdash \exists {v} \in {\bf Q\/} ({v}\in L) \wedge \exists {w} \in {\bf Q\/} ({w}\in U).$ \\
 This requires showing that there is an open cover $\{W_i\}$ such that for each index $i$, there is a locally constant function over $W_i$ in $L(W_i)$ and a locally constant function over $W_i$ in $U(W_i)$. This follows because for all $W_i \subseteq W$, the function $v_i:W_i\rightarrow {\bf Q}$ with $v_i(x)<0$ for all $x$ belongs to $L(W_i)$. Likewise, the function $w_i:W_i\rightarrow {\bf Q}$ with $w_i(x)>1$ for all $x$ belongs to $U(W_i)$.

  \item[2. Unbounded.] $W \Vdash \forall {v}, {w} \in {\bf Q\/} (v<w \wedge {w}\in L \Rightarrow {v}\in L) \wedge (v<w \wedge {v}\in U \Rightarrow {w}\in U)$. \\
  This follows from monotonicity (Lemma~1 above), and the construction of $L$ and $U$ to include all negative rational numbers in $L$, and rational numbers greater than one in $U$ respectively. To see this, for any $W^\prime \subseteq W$,
  consider first $v\in {\bf Q\/}(W^\prime)$ and $w\in {\bf Q\/}(W^\prime)$ with $v(x) \in [0,1)$ and $v<w\leq 1$.
  If $w \in L$, from monotonicity we have, $vp+({1}-v)r \prec q$. Hence, $v\in L$. If $v$ has $v(x)<0$ for all $x$, it is in $L$ by definition.
  The second half of the statement follows in a similar manner.

  \item[3. Open.] $W \Vdash \forall v \in {\bf Q\/} ({v}\in L \Rightarrow (\exists w \in {\bf Q}) (w>v \wedge w\in L))$ and \\
  $W \Vdash \forall {v} \in {\bf Q\/} ({v}\in U \Rightarrow (\exists w \in {\bf Q}) (w<v \wedge {w}\in U)).$

  For the first statement, it must be shown that for a locally constant function $v:W^\prime \rightarrow \mathbb{Q}$ defined on
  $W^\prime \subseteq W$ with $v\in L(W^\prime)$ there is an open cover $\{ W_i^\prime \}$ of $W^\prime$ and locally constant functions
  $w_i:W_i^\prime \rightarrow \mathbb{Q}$ such that $w_i(x)>v(x)$ for all $x\in W_i^\prime$ and ${w_i}\in L(W_i^\prime))$.

  If, for some $W_i^\prime$, it is the case that $v(x)<0$ for all $x \in W_i^\prime$ the result follows because for any $v<0$ in $\bf {Q}$ there is a locally constant function $w_i:W_i^\prime \rightarrow \mathbb{Q}$ such that $w_i(x)>v(x)$ for $x\in W_i^\prime$ and $w_i(x)<0$.

  For no $W_i^\prime$ and $v\in L(W_i^\prime)$ is it the case that $v(x)=1$ for $x \in W_i^\prime$, because this would imply $p \precsim q$, contrary to assumption. So, consider $v$ with $0\leq v(x) <1$. When $v=0$, we have $W_i^\prime \Vdash r\prec q \precsim p$ from (5). From independence, take $a: W_i^\prime \rightarrow (0,1]$ such that $W_i^\prime \Vdash r \prec ar+(1-a)q \prec q \precsim p$.
  It follows that $W_i^\prime \Vdash r \prec ar+(1-a)q \prec p$. By continuity, there is an open cover $\{U_j\}$ of $W_i^\prime$ and {\it rational-valued\/} continuous functions  $b_j: U_j\rightarrow (0,1)\cap \mathbb{Q}$ such that $U_j \Vdash r \prec b_jp+(1-b_j)r \prec ar+(1-a)q \prec q$ (specifically, for $b_j>0$). Collating, there is $w_i:W_i^\prime \rightarrow \mathbb{Q}$ such that $w_i(x)>v(x)=0$ for $x\in W_i^\prime$ and ${w_i}\in L(W_i^\prime))$.

  The remaining case is $v$ with $v(x)>0$ for all $x$. From claim~1 (monotonicity) and $v\in L(W^\prime)$ we have $W^\prime \Vdash r \prec vp + (1-v)r \prec q$. From independence,
  \[ W^\prime \Vdash r \prec vp + (1-v)r \prec a_i [vp+(1-v)r]+(1-a_i)q \prec q \precsim p. \]
  From continuity, there is $\{ W_i^\prime \}$ such that $\bigcup_i W_i^\prime = W^\prime$ and rational-valued continuous functions $b_i:W_i^\prime \rightarrow (0,1)\cap \mathbb{Q}$ such that
  \[ W_i^\prime \Vdash vp+(1-v)r \prec b_ip + (1-b_i)[vp+(1-v)r] \prec a_i [vp+(1-v)r]+(1-a_i)q \prec q.\]
  Upon setting $w_i = v+b_i(1-v)$ this yields
  \[W_i^\prime \Vdash vp+(1-v)r \prec w_ip + (1-w_i)r \prec q.\]
  So, $w_i(x)>v(x)$ for all $x\in W_i^\prime$ and $w_i \in L(W_i^\prime)$.

  The case of $U(W)$ is analogous.

  \item[4. Arbitrarily close.] $W \Vdash \forall v, w \in {\bf Q\/} (v<w \Rightarrow v\in L \vee w\in U)$ \\
  To show this, we need to show that there is an open cover $\{W_i\}$ of $W$
  such that for each $i$, and all $x\in W_i$, $(v(x)p(x)+({1}-v(x))r(x) \prec q(x))$ or $q(x) \prec (w(x)p(x)+({1}-w(x))r(x))$. This follows from negative transitivity, noting that, over $W$, $(vp+(1-v)r \prec (wp+(1-w)r$ (from monotonicity, Claim~1). Then there exists an open cover $\{W_i\}$ of $W$ such that for each $i$ and all $x\in W_i$ either $(v(x)p(x)+({1}-v(x))r(x) \prec q(x))$ or $q(x) \prec (w(x)p(x)+({1}-w(x))r(x))$.

  \item[5. Disjoint.] $W \Vdash (\forall v \in {\bf Q\/}) \neg (v\in L \wedge v\in U)$\\
  This follows from the asymmetry of $\prec$: $vp+(1-v)r \prec q \Rightarrow \neg (q \prec vp+(1-v)r)$ and $q \prec vp+(1-v)r \Rightarrow \neg (vp+(1-v)r) \prec q$.
\end{description}

Now writing $\hat{v}$ for the constant function on $X$ with value $v\in \mathbb{Q}$, consider for a point $x\in W$ the following sets of rationals:
\[ L_x = \{ v\in \mathbb{Q} | \exists \mbox{open} \  V \subseteq W: x\in V \mbox{and} \ \hat{v}|V \in L(V) \}, \]
\[ U_x = \{ w\in \mathbb{Q} | \exists \mbox{open} \  V \subseteq W: x\in V \mbox{and} \ \hat{w}|V \in U(V) \}. \]
From the properties of $L(V)$ and $U(V)$ it follows that $L_x$ and $U_x$ (as subsets of $\mathbb{Q}$) form a Dedekind cut (satisfy the Dedekind conditions). There is a {\em unique\/} real number $\sup L_x$ =$\inf U_x$ in $\mathbb{R}$ which corresponds to the cut $(L_x, U_x)$. The function
\[a^*: W \rightarrow \mathbb{R} \]
defined by $a^*(x) = \sup L_x$ has, for $v, w \in \mathbb{Q}$,
\begin{equation}
v < a^*(x) < w \quad \mbox{iff} \ v\in L_x \ \mbox{and} \ w\in U_x.
\end{equation}
It follows from (7) that $a^*$ is continuous. The following proof of continuity follows \citep[][p.~324]{maclane1996} and is included for completeness. If $(v,w)$ is a rational interval and $x\in (a^{*})^{-1}(v,w)$, then by (7) and the definition of $L_x$ and $U_x$, there are neighborhoods $V$ and $V^\prime$ of $x$ such that $\hat{v}|V \in L(V)$ and $\hat{w}|V^{\prime} \in U(V^\prime)$. Then for any point $y\in V\cup V^\prime$, again by (7), $y\in  (a^{*})^{-1}(v,w)$. Thus $V\cup V^\prime \subseteq (a^{*})^{-1}(v,w)$, and $(a^{*})^{-1}(v,w)$ is an open subset of $W$.

Finally, it follows that $W \Vdash q\sim a^*p+({1}-a^*)r$. Suppose that for some set $W^\prime \subseteq W$ with $W^\prime \Vdash a^*p+({1}-a^*)r \prec q$.
Then, by continuity, there is an open cover $\{W_i^\prime\}$ of $W^\prime$ and continuous rational functions  $b_i: W_i^\prime \rightarrow (0,1)\cap \mathbb{Q}$ such that for all $x\in W_i^\prime$
  $a^*(x)p(x)+(1-a^*(x))r(x) \prec b_i(x)p(x) + (1-b_i(x))(a^*(x)p(x)+(1-a^*(x))r(x)) \prec q(x)$. This contradicts $a^*(x) = \sup L_x$, implying $W \Vdash \neg(a^*p+({1}-a^*)r \prec q)$.
  By an analogous argument $W \Vdash \neg(q \prec a^*p+({1}-a^*)r)$.
  In other words, $W \Vdash q\sim a^*p+({1}-a^*)r$. This proves the claim. $\blacksquare$

 The next claim extends independence for weak preference ($\precsim$) and indifference ($\sim$). The new Assumption~4 plays a role here---it is asserted that one of the two conditions referenced there holds, although not that they collectively exhaust all the possibilities. The proposition just claims that if one or the other conditions can be asserted to hold over the open set $W$, then independence of $\precsim$   and $\sim$ follows.

  \begin{claim}[Independence for $\precsim$ and $\sim$]
  Suppose $p,q \in P(W)$, and $a:W\rightarrow  [0,1]$ is a continuous function. Suppose also that one of the following two conditions holds:
  (I)~$W \Vdash [\forall f, g \in P(W), f \sim g]$, or (II)~$W \Vdash [\exists f, g \in P(W), f \prec g]$.
  Then, for all $r\in P(W)$,
  \begin{enumerate}
  \item $W \Vdash p \sim q \Rightarrow ap+(1-a)r \sim aq+(1-a)r,$
  \item $W \Vdash p \precsim q \Rightarrow ap+(1-a)r \precsim aq+(1-a)r.$
  \end{enumerate}
  \end{claim}
\noindent {\it Proof.\/}
Suppose first that $W \Vdash p\sim q$. If it is the case that (I)~$W \Vdash [\forall f, g \in P(W), f \sim g],$ the claim follows. In this case, for all $W^\prime \subseteq W$, and $p, q \in P(W^\prime)$, $p \sim q$.
For  $p, q \in P(W)$, $p \sim q \Rightarrow ap+(1-a)r \sim aq+(1-a)r$ because $ap+(1-a)r \in P(W)$ and $aq+(1-a)r \in P(W)$.
Now suppose that $W \Vdash [\exists f, g \in P(W), f \prec g]$.
From $f \prec g$ and negative transitivity there is an open cover $\{ V_j\}$ of $W^\prime$ such that for each index $j$,
$V_j \Vdash p \sim q \prec g$ or $f \prec p \sim q$. For $r\in P(W)$,
suppose $V_j \Vdash aq+(1-a)r \prec ap+(1-a)r$ and consider the case where $V_j \Vdash p\sim q \prec g$ (the argument for the  $V_j \Vdash f \prec p\sim q$ case is similar).
From Assumption 2, for all $b: V_j \rightarrow (0,1]$, $p \sim q \prec bg + (1-b)q$, and
\[ ap + (1-a)r \prec a(bg + (1-b)q) + (1-a)r. \]
Since by assumption, $aq+(1-a)r \prec ap+(1-a)r$, it follows that:
\[ aq+(1-a)r \prec ap+(1-a)r \prec a(bg + (1-b)q) + (1-a)r.\]
For any $b$, from Assumption 3, there exists an open cover $\{U_k\}$ of $V_j$, and continuous functions $a^*_k:U_k \rightarrow (0,1)$ such that:
\[U_k \Vdash a_k^*[a(bg+(1-b)q)+(1-a)r]+(1-a_k^*)[aq+(1-a)r] \prec ap+(1-a)r. \]
The term on the left can be simplified to $a[a_k^*bg + (1-a_k^*b)q] + (1-a)r.$ Since $p\prec [a_k^*bg + (1-a_k^*b)q]$,
\[ap+(1-a)r \prec a_k^*[a(bg+(1-b)q)+(1-a)r]+(1-a_k^*)[aq+(1-a)r].\]
So, $ap+(1-a)r\prec ap+(1-a)r$, violating asymmetry. Since $U_k \subseteq W$,
\[W \Vdash \neg (aq+(1-a)r \prec ap+(1-a)r).\]
In other words, $W \Vdash ap+(1-a)r \precsim aq+(1-a)r.$ In a similar manner, one can deduce $\neg (ap+(1-a)r \prec aq+(1-a)r)$, or $aq+(1-a)r \precsim ap+(1-a)r.$ It follows that $aq+(1-a)r \sim ap+(1-a)r.$
This completes the proof of the first part of the claim. The proof for the second part is similar. $\blacksquare$

\begin{claim}[Local representation by affine mapping]
Suppose, for $W \in \mathcal{O}(X)$, $ W \Vdash [\forall p, q \in P(W) p \sim q]$ or  $W \Vdash [\exists p, q \in P(W), p \prec q].$ Then there is a mapping $f_W$ which represents preferences over $W$ in the sense that, for $r, s\in P(W)$,
\[ W \Vdash r \prec s \quad \Longleftrightarrow \quad f_W(r) < f_W(s).\]
Moreover, for $ar+(1-a)s \in P(W)$ and $a:W\rightarrow [0,1]$ continuous, $f_W$ satisfies $f_W(ar+(1-a)s) = af_W(r)+(1-a)f_W(s)$. Furthermore, if $V\subseteq W$, then $f_V$ similarly represents preferences over the restriction $P(V)$ of $P(W)$.
If $\rho_{WV}$ denotes the restriction of $P(W)$ to $P(V)$ and $\rho^\prime_{WV}$ the restriction of ${\bf R}(W)$ to ${\bf R}(V)$, we have $f_V\circ \rho_{WV} = \rho^\prime_{WV} \circ f_W $.
\end{claim}
\noindent {\it Proof.\/}
The first case is of indifference between all variable lotteries: $W \Vdash [\forall p, q \in P(W) p \sim q]$ (i.e., for all $V \subseteq W$ and all $p, q \in P(V)$, $p\sim q$).
Then $f_W:P(W) \rightarrow {\bf R}(W)$ must be chosen such that $f_W(r)$ is the same continuous function in ${\bf R}(W)$ for all $r \in P(W)$. Additionally, for any $V \subseteq W$ and
all $r \in P(V)$, $f_{V}(r)$ must be the restriction of $f_W(r)$ to $V$ (i.e., $f_W(r)|_V \in {\bf R}(V)$).
Note that, since the restriction operation is order-preserving (Assumption~1), when $p|_W \sim q|_W$ is asserted, it is also the case that $p|_V \sim q|_V$.
Since $f_W(p)=f_W(q)$, it follows that $f_V(p|_V)=f_V(q|_V)$. However, $P(V)$ may contain
new variable lotteries that are not restrictions of variable lotteries over $W$ (as was seen in Example~2). These new lotteries must also map to the same $f_V(r)$. For $V \subseteq W$, let $\rho_{WV}(r)$ be the
restriction of $r$ to $V$. Then $\rho_{WV}$ is mapping from $P(W)$ to $P(V)$. If restriction from ${\bf  R}(W)$ to ${\bf  R}(V)$ is denoted by $\rho^\prime_{WV}$, it follows that  $f_V\circ \rho_{WV} = \rho^\prime_{WV} \circ f_W$.

The case of $W \Vdash [\exists p, q \in P(W), p \prec q]$ is considered next. Assume there are $p, q \in P(W)$ with $W \Vdash p \prec q$, and consider the set
$ [pq] \equiv \{r: W\rightarrow \mathbb{S}^n | p \precsim r \precsim q \}$.
For $r \in [pq]$, and $a^* \in {\bf R}(W)$ such that $W \Vdash a^* q+(1-a^*)p \sim r$, define $f_W(r) = a^*$.
Such a continuous function $a^*$ must exist from Lemma~2. It follows that $f_W(q)=1$ and $f_W(p)=0$.
For $V \subseteq W$ the restriction $\rho_{WV}: P(W) \rightarrow P(V)$ has been defined to be order-preserving. So also is $\rho^\prime_{WV}$, the restriction from ${\bf R}(W)$ to ${\bf R}(V)$, order-preserving in $<$.\footnote{Note
that $P(V)$ and ${\bf R}(V)$ may have new elements (i.e., elements not arising as restrictions of elements of $P(W)$ and ${\bf R}(W)$) that are also now ranked.}
It is now shown that $f_W$ is also order-preserving, and that rankings on restrictions are related.

From monotonicity (Lemma~1), for any $W$,
\[ W \Vdash f_W(r) < f_W(s) \quad \mbox{iff} \quad f_W(r)q + (1-f_W(r))p \prec f_W(s)q + (1-f_W(s))p. \]
For all $r, s \in [pq]$ and given continuous $a:W\rightarrow \mathbb{R}$, if $W\Vdash r \sim f_W(r)q + (1-f_W(r))p,$ and
$W\Vdash s \sim f_W(s)q + (1-f_W(s))p,$ then (by two applications of Claim~3)
\[ W \Vdash ar+(1-a)s \sim a[ f_W(r)q + (1-f_W(r))p] + (1-a)[f_W(s)q + (1-f_W(s))p]. \]
This simplifies to,
\[ W \Vdash ar+(1-a)s \sim [af_W(r)+(1-a)f_W(s)]q + [1- (af_W(s)+(1-a)f_W(s))]p ,\]
and so, $f_W(ar+(1-a)s) = af_W(r)+(1-a)f_W(s)$. In words, $f_W$ is an order-preserving affine function. Clearly, if $f_W$ represents preferences, so
must any positive linear transformation $af_W+b$ with $a>0$.
By the same logic, $f_V$ is order-preserving and affine over $[pq]$ restricted to $V$ (i.e., $[p|_Vq|_V]$ ).
Here again, for $V \subseteq W$, it must be the case that $f_V\circ \rho_{WV} =  \rho^\prime_{WV}\circ f_W$ (so long as the same $p$ and $q$ are used for calibration).
This is because $W \Vdash a^* q+(1-a^*)p \sim r$
implies $V \Vdash a^* q+(1-a^*)p \sim r$. With $f_V(q)=1$ and $f_V(p)=0$, $f_V(r)$ is just the restriction of $a^*$ to $V$.

Next, the representation on $[pq]$ is extended to the rest of $P(W)$. If there are $r$ and $s$ in $P(W)$ such that $[pq] \subset [rs]$ a familiar argument
\citep[e.g.,][p.~114]{fishburn1970} extends to the case of functions. Find an $f$ that represents preferences over $[rs]$ (such an $f$ exists by Lemma~2), ensuring that $f(p)=0$ and $f(q)=1$. This is always possible: for instance, if $f^\prime$ represents preferences over $[rs]$ with $f^\prime(r)=0$ and $f^\prime(s)=1$, take
$f(\cdot) = (f^\prime(\cdot)-f^\prime(p))/(f^\prime(q)-f^\prime(p))$ ($f^\prime$ is a positive linear transformation of $f$).
Now if $s\prec p \prec q$, there is a continuous real function $a$ such that $f(p) = af(q)+(1-a)f(s)$, so that $f(s) = -a/(1-a)$. If $p\prec q \prec r$,
there is a continuous real function $b$ such that
$f(q) = bf(r)+(1-b)f(p)$ or $f(r)=1/b$. If $p\prec r \prec q$, there is a continuous real function $c$ such that $f(r) = cf(q)+(1-c)f(p)$ or $f(r)=c$. If there is some
other $[r^\prime s^\prime]$ that contains $[pq]$, and another function $f^*$ that represents preferences, then from the uniqueness of the representation in Lemma~2,
$f(t)=f^*(t)$ for all $t\in [rs]\cap [r^\prime s^\prime]$.

The remaining case to consider is of $r\in P(W)$ that are sometimes uncomparable with $p$ or $q$ (e.g., if preference for $r$ changes relative to $p$ and $q$, as in Figure~2).
This is addressed next. Given $p\prec q$, and $r\in P(W)$, it follows from negative transitivity that there is an open cover $\{W_i \}$ of $W$ such that $W_i \Vdash (r\prec q)\vee (p \prec r)$. Consider the case of $r\prec q$ (the case of $p\prec r$ is similar). From independence there are real-valued functions $a$ and $b$ such that $r \prec aq+(1-a)r\prec q$ and $p \prec bq+(1-b)p \prec q$. Starting with $p \prec bq+(1-b)p$, and given $aq+(1-a)r$, from another application of negative transitivity, if follows that: for an open cover $\{ W^j_i\}$ of $W_i$, $W^j_i \Vdash p \prec aq+(1-a)r \ \mbox{or} \ aq+(1-a)r \prec bq+(1-b)p$. In the first case, $p\prec aq+(1-a)r\prec q$ and $r \prec aq+(1-a)r \prec q$ over $W_i^j$. In the second case,
$r \prec bq+(1-b)p\prec q$ and $p\prec bq+(1-b)p \prec q$. So one can assert, $ W_i^j \Vdash (\exists p^\prime) (p \prec p^\prime \prec q) \wedge (r \prec p^\prime \prec q)$.
From the above, $f_1$ can be constructed with domain $[pq]$ and $f_1(q)=1$, $f_1(p^\prime)=0$. In this case, if $p^\prime \sim \gamma q+(1-\gamma)p$,
$f_1(p) = -\gamma/(1-\gamma)$. Similarly, $f_2$ can be constructed with domain $[rq]$ and $f_2(q)=1$, $f_2(p^\prime)=0$. In this case, if $p^\prime \sim \delta q+(1-\delta)r$,
$f_2(r) = -\delta/(1-\delta)$. It remains to be shown that for all $t \in [pq] \cap [rq]$, $f_1(t) = f_2(t)$.

Let $D = [pq] \cap [rq]$ with domain of definition $W_i^j$. The mappings $f_1$ and $f_2$ are defined on $D$ and are order-preserving, with $f_1(q)=f_2(q)=1$ and $f_1(p^\prime)=f_2(p^\prime)=0$. \footnote{The functions $f_1$ and $f_2$ should be indexed by $W_i^j$, but this becomes cumbersome.}
Since $f_1$ and $f_2$ both represent preferences over $D$, and uniqueness up to positive linear transformations (see Lemma 6 below),
$f_2 = \alpha f_1 + \beta$, where $\alpha$ and $\beta$ are continuous real-valued functions and $\alpha>0$. Since $f_2(p^\prime) = \alpha f_1(p^\prime) + \beta$,
$\beta = 0$. Since $f_2(q) = \alpha f_1(q) + \beta$, $\alpha = 1$. Hence, for all $t\in D$, $f_1(t) = f_2(t)$. This result continues to hold for restrictions to $V \subset W_i^j$
because restriction is order-preserving (the relative rankings of $p, q, r,$ and $p^\prime$ are preserved). For $t \in [p|_Vq|_V] \cap [r|_Vq|_V]$, $f_1(t) = f_2(t)$ (again, the
index $V$ is dropped from the mappings).

Both $f_1$ and $f_2$ are now re-scaled to make $q$ the unit, and $p$ the zero:
\[ f_1^*(\cdot) = \frac{f_1(\cdot) - f_1(p)}{f_1(q) - f_1(p)} \quad \quad f_2^*(\cdot) = \frac{f_2(\cdot) - f_1(p)}{f_1(q) - f_1(p)}. \]
Now $f_1^*(q) = f_2^*(q) = 1$ (because $f_1(q) = f_2(q) = 1$) and $f_1^*(p)=0$. For every $V\subseteq W_i^j$ and $t \in [p|_Vq|_V]\cap [r|_Vq|_V]$ it
is readily checked that $f_1^*(t) = f_2^*(t)$.
For any other $t$ in $[p|_Vq|_V]\cup [r|_Vq|_V]$, $f_1^*$ or $f_2^*$ are used for valuation as relevant. In this manner, any distribution in $P(W_i^j)$ can be assigned a value.
Denote this mapping from $P(W_i^j)$ to ${\bf R}(W_i^j)$ by $f_{W_i^j}$.
An explicit calculation for $r$ is as follows for illustration.
Since $f_1(p) = -\gamma/(1-\gamma)$, $f_1(q)=1$, and $f_2(r) = -\delta/(1-\delta)$:
\[ f_{W_i^j}(r) = f_2^*(r) = \frac{\gamma-\delta}{1-\delta}.\]
If, over any open subset $V$, $r \prec p$, it follows that $\gamma<\delta$ and $f_2^*(r)<0$. If  $p \prec r$, it follows that $\delta<\gamma$ and $f_2^*(r)>0$.

The representing mappings $f_{W_i^j}$ can be collated to a unique mapping over all of $W$. For each $W_i^j$ there is
a function $f_{W_i^j}$ representing preferences over $W_i^j$. Now consider two such sets $W_i^j$ and $W_i^k$, with $V = W_i^j \cap W_i^k$.
Since $p$ and $q$ are defined over all of $W$, $f_{W_i^j}(p|_{W_i^j})=f_{W_i^k}(p|_{W_i^k})=0$ and
$f_{W_i^j}(q|_{W_i^j})=f_{W_i^k}(q|_{W_i^k})=1$. For any $t \in P(W_i^j \cup W_i^k)$, suppose $f_{W_i^j}(t|_{W_i^j})$ and $f_{W_i^k}(t|_{W_i^k})$ are defined and
$f_{W_i^j}(t|_{W_i^j})|_V=f_{W_i^k}(t|_{W_i^k})|_V$. Then $f_{W_i^j}(t|_{W_i^j})$ and $f_{W_i^k}(t|_{W_i^k})$ can be glued together to yield a continuous real
function with domain $W_i^j \cup W_i^k$ (i.e., $f_{W_i^j \cup W_i^k}(t)$). Such functions represent preferences for variable lotteries over $W_i^j \cup W_i^k$.
The same holds over all such overlapping sets $V$, and the representing mappings can thus be
collated to yield a unique mapping $f_{W_i}$ over $W_i$. It was earlier asserted that $W_i \Vdash (r\prec q)\vee (p \prec r)$,
and the case of $r\prec q$ was considered. An identical argument holds for all the $W_i$ with $r\prec q$. The case of $p\prec r$ similarly leads to representing functions $f_{W_i}$ with $f_{W_i}(q)=1$ and $f_{W_i}(p)=0$. Once again, with $V = W_i\cap W_j$, if $ f_{W_i}(t|_{W_i})|_V=f_{W_j}(t|_{W_j})|_V$ the functions can be uniquely collated over $W$. The claim follows. Finally, note $f_V\circ \rho_{WV} =  \rho^\prime_{WV}\circ f_W$ holds here as well for the same reasons as before (restriction is order-preserving, the representing mappings are order-preserving, and the same $p$ and $q$ are used as the zero and unit values of $f$). $\blacksquare$

\begin{claim}[Local expected utility representation] Given an affine function $f$ that represents preferences, let $u_i = f(\delta_i)$ for each $z_i\in Z(W)$. Then, for $p, q \in P(W)$ and $p_i (\mbox{respectively} \  q_i)$ denoting the $i$-th coordinate of $p (\mbox{respectively} \ q)$,
\[ W \Vdash p \prec q \quad \mbox{if and only if} \quad \sum_{i=1}^n p_iu_i < \sum_{i=1}^n q_iu_i .\]
\end{claim}
\noindent {\it Proof.\/}
For $z_i\in Z(W)$, define $u_i = f(\delta_i).$  This makes $u_i(\cdot)$ a continuous function from $W$ to $\mathbb{R}$. The expected utility representation:
$f(p) = \sum_{i=1}^n p_i(x)u_i(x)$ is now deduced. Note that any $p\in P(W)$ can be written as $p = p_1 \delta_1+ p_2\delta_2+ \ldots p_n\delta_n$.
If $p=\delta_n$ (i.e., $p_n=1$) at any $x\in W$, then $f(p(x))=f(\delta_n(x))$ and the representation holds at $x\in W$. In fact, the representation holds for all
$y\in W$ as well. For any $y$ with $p_n(y)<1$,  $1-p_n(y)>0$ and
\[ p = \left( \frac{p_1}{1-p_n}\delta_1 + \ldots  +\frac{p_{n-1}}{1-p_n}\delta_{n-1} \right)(1-p_n) + p_n\delta_n. \]
Denote $q =  \left( \frac{p_1}{1-p_n}\delta_1 + \ldots  +\frac{p_{n-1}}{1-p_n}\delta_{n-1} \right)$ and $q_i =  \frac{p_i}{1-p_n}$.
Clearly, $q:W\rightarrow \mathbb{S}^n$.  Since $f$ is affine, ir follows that
$f(p) = (1-p_n)f(q) + p_nf(\delta_n).$ Now consider $q$ and suppose $q_{n-1}=1$ at some $x$. Then $q(x)=\delta_{n-1}(x)$ and $f(q) = f(\delta_{n-1})$ at $x$. It follows that
$f(p) = (1-p_n)f(\delta_{n-1}) + p_nf(\delta_n)$. So again, the representation holds at $x$. When $q_{n-1}<1$,
\[ q  = \left( \frac{q_1}{1-q_{n-1}}\delta_1 + \ldots  +\frac{q_{n-2}}{1-q_{n-1}}\delta_{n-2} \right)(1-q_{n-1}) + q_{n-1}\delta_{n-1}. \]
Denoting $ q^\prime  = \left( \frac{q_1}{1-q_{n-1}}\delta_1 + \ldots  +\frac{q_{n-2}}{1-q_{n-1}}\delta_{n-2} \right),$ it is again the case that
$q^\prime:W\rightarrow \mathbb{S}^n$.
Then it follows that $f(q) = (1-q_{n-1})f(q^\prime) + q_{n-1}f(\delta_{n-1}).$
Upon substitution,
$$f(p) = (1-p_n)(1-q_{n-1})f(q^\prime) + p_{n-1}f(\delta_{n-1})+p_nf(\delta_n).$$
Repeated applications of this step yields $$f(p) = \sum_i p_if(\delta_i).$$
It follows that, for all $W \subseteq X$,
\[ W \Vdash p \prec q \quad \mbox{if and only if} \quad \sum_{i=1}^n p_iu_i < \sum_{i=1}^n p_iu_i. \quad \blacksquare\]

Following convention, one may use $u(\cdot)$ instead of $f(\cdot)$ as the notation for the representing mapping from $P(W)$ to ${\bf R}(W)$. The lemma shows that
$u(p) = \sum_{i=1}^n p_iu_i$, where $u_i = u(\delta_i)$.

\begin{claim}[Uniqueness]
Any function $u(\cdot)$ representing preferences over some $W\subseteq X$ is unique up to positive linear transformations.
\end{claim}
\noindent {\it Proof.\/}
Suppose $u(\cdot)$ is an affine mapping representing preferences over $P(W)$ for some $W \subseteq X$. For uniqueness up to positive linear transformations, note that if $v(\cdot) = au(\cdot) + b$ (for continuous real functions
$a, b: W\rightarrow \mathbb{R}$ with $a>0$) then $v$ represents preferences as well. Conversely, suppose $v(\cdot)$ and $w(\cdot)$ are two functions that represent preferences.
If the first condition in Assumption~4 holds, i.e., $W \Vdash \forall f, g \in P(W), f \sim g$, then $v(\cdot) = \bar{v}$ and $w(\cdot) = \bar{w}$. In this case, $w(\cdot)$  can clearly
be written as a positive linear transformation of $v(\cdot)$. Now consider the second condition: $W \Vdash \exists f, g \in P(W), f \prec g$.
For $r\in P(W)$ define (for fixed $v(f)$, $v(g)$, $w(f)$, and $w(g)$):
\[ f_1(r) = \frac{v(r)-v(f)}{v(g)-v(f)} \quad \mbox{and} \quad f_2(r) = \frac{w(r)-v(f)}{w(g)-w(f)}.\]
This requires $v(g)-v(f)>0$ and $w(g)-w(f)>0$ for all $x$, which follows from the assumption that $f \prec g$ and the representation result.
The functions $f_1$ and $f_2$ are positive linear transformations of $v(\cdot)$ and $w(\cdot)$ respectively, and so also represent preferences. If $r=f$ then $f_1(r)=f_2(r)=0$
and if $r=g$ then $f_1(r)=f_2(r)=1$. For any $r \sim ag + (1-a)f$ we have $f_i(r) = a f_i(g)+(1-a)f_i(f)$ for
$i=1,2$. It follows that for all $r$, $f_1(r)=f_2(r)$, and so,
\[ w(r) = \frac{w(g)-w(f)}{v(g)-v(f)}v(r) + w(f) - v(f)\frac{w(g)-w(f)}{v(g)-v(f)}.\]
So $w$ must be a positive linear transformation of $v$. $\blacksquare$

Most of the work is now done. It remains to invoke Assumption~4.
\begin{proof}[Proof of Theorem 1]
It needs to be shown that there is an open cover $\{W_i\}$ with $\bigcup_i W_i = X$, and functions $u_{W_i}:P(W_i) \rightarrow {\bf R}(W_i)$, such that, for each $W_i$,
\begin{equation}
W_i \Vdash p|_{W_i} \prec  q|_{W_i} \Longleftrightarrow
\sum_{z=1}^n u_{W_i}(\delta_z|_{W_i})p_z|_{W_i} < \sum_{z=1}^n u_{W_i}(\delta_z|_{W_i})q_z|_{W_i}.
\end{equation}
Assumption~4 states that there is an open cover $\{ W_i\}$ of $X$ and for each $W_i$ one of the conditions of the assumption holds. I.e.,
\begin{quote}
$ W_i \Vdash [ \forall p, q \in P(W_i), p \sim q] \ \mbox{or} \  [\exists p, q \in P(W_i), p \prec q].$
\end{quote}
From Lemma~4, in each of the two cases, there is an affine $u_{W_i}: P(W_i) \rightarrow {\bf R}(W_i)$ that represents preferences in the sense of Equation~8.
Lemma~5 yields the local expected utility representation over $W_i$. Lemma~6 provides the uniqueness result over $W_i$. The fact that the $W_i$ cover all of $X$ ensures
the result. Lemma~4 also provides the result on restrictions to $V\subseteq W$: if $u_W$ exists and represents preferences over $W$ and, for $V \subseteq W$, $u_V$ represents
preferences over $V$, and $\rho_{WV}$ is the restriction of $P(W)$ to $P(V)$ and $\rho_{WV}^\prime$ is the restriction of ${\bf R}(W)$ to ${\bf R}(V)$, then
$u_V \circ \rho_{WV}  =  \rho_{WV}^\prime \circ u_W $.
\end{proof}

\begin{proof}[Proof of Corollary 1]
$X$ can be partitioned into two disjoint sets $W^1$ and $W^2$ such that the first (indifference) condition holds over $W^1$ and
the second (existence of two ranked variable lotteries) condition holds over $W^2$. Without loss of generality one can
consider separate coverings of $W^1$ and $W^2$. The indifference case is exactly as in Theorem~1. For the other case,
by Assumption~6, there exists a covering $\{W_i\}$ of $W^2$, and two compatible families $\{p_i\}$ and $\{q_i\}$ collating to
$p:W^2 \rightarrow \mathbb{S}^n$ and $q:W^2 \rightarrow \mathbb{S}^n$  respectively, satisfying $W_i \Vdash p_i \prec q_i$ for all $i$.
Clearly, Assumption~6 implies that Assumption~4 is satisfied. Hence, there are affine functions $u_{W_i}$ such that Equation~1 is satisfied. For every $i$, from the uniqueness result (Lemma~6), one can set $u_{W_i}(p_i) = 0$ and  $u_{W_i}(q_i) = 1$. Further,
$\{u_{W_i}(p_i)\}$ and $\{ u_{W_i}(q_i)\}$ are compatible families (equal to constant functions with values 0 and 1 respectively).
Consider any two $W_i$ and $W_j$ with $W_i \cap W_j \neq \emptyset$.
With $u_{W_i}(p_i) = u_{W_j}(p_j)=0$ and $u_{W_i}(q_i) = u_{W_j}(q_j)=1$, it follows that for every
$r|_{W_i\cup W_j} \in P(W_i \cup W_j)$, $u_{W_i}(r|_{W_i}) = u_{W_j}(r|_{W_j})$ over $W_i \cap W_j$.
So, there is a unique $u_{W_i \cup W_j}(r|_{W_i\cup W_j})\in {\bf R}(W_i\cup W_j)$ with restriction equal to $u_{W_i}(r|_{W_i})$ over $W_i$ and
 $u_{W_j}(r|_{W_j})$ over $W_j$. In general, for $W^2 \in \mathcal{O}(X)$ and an open cover $\{W_i\}$ of $W^2$,
consider a pair of compatible families $\{r_i\}$ and $\{s_i\}$ collating to $r$ and $s$ in $P(W^2)$ respectively.
There exist $u_{W_i}(r_i), u_{W_i}(s_i) \in {\bf R}(W_i)$ for all $i$, also agreeing on overlaps. These can be
collated to mappings $u_{W^2}(r)$ and $u_{W^2}(s)$ respectively.
For $V \subseteq W^2$,
\[ V \Vdash r|_V \prec s|_V \Longleftrightarrow u_W(r)|_V <  u_W(s)|_V.\]
A global mapping $u_X \equiv u$ can be defined over all of $X$ to represents preferences for lotteries in $P(X)$ --
since $W^1$ and $W^2$ are disjoint, pasting functions over $W^1$ and over $W^2$ is straightforward (there is no intersection
on which the functions must match).
For lotteries that appear only for $W \subset X$ as new elements of $P(W)$ one may still select $p$ as the zero and $q$ as the unit to calibrate the lottery with domain restricted to $W$.

Part~2 of the proof is immediate from the proof of Lemma~4.
\end{proof}

\begin{proof}[Proof of Theorem 2]
For sufficiency, observe that Assumption~5 implies that Assumption~4 holds. This is because $[ (\forall p, q \in P(X))p\sim q]$ is the
negation of $[ (\exists p, q \in P(X))p\prec q]$.
Let $ \llbracket \phi \rrbracket \in \mathcal{O}(X)$ denote the truth value of proposition $\phi$ (i.e., the stages where it holds).
Then, by Assumption~5, $ \llbracket (\forall p, q \in P(X))p\sim q \rrbracket \cup \llbracket (\exists p, q \in P(X))p\prec q \rrbracket = X$.
The conclusion of Theorem 1 then follows.

Next, it will be shown that there is a mapping $u: P(X) \rightarrow {\bf R}(X)$ such that, for all $r\in P(X)$, $u_{W_i}(r|_{W_i}) = u(r)|_{W_i}$.
From Assumption~4 (which is an implication of Assumption~5)
there is a covering $\{W_i\}$ of $X$ such that either there are $p_i, q_i \in P(W_i)$ with $p_i \prec q_i$ or, for all $W_i^\prime \subseteq W_i$ and
$p_i, q_i \in P(W_i^\prime)$, $p_i \sim q_i$. If the latter, let $u(p_i)=u(q_i)=\bar{u}$ for all $p_i, q_i \in P(W_i)$, where $\bar{u}$ is a continuous
real-valued function . If there is another $W_j$ with $p_j \sim q_j$ for all
$p_j, q_j \in P(W_j)$, and $W_i\cap W_j \neq \emptyset$, the function chosen over $W_j$ must coincide with $\bar{u}$ over $W_i\cap W_j$,
so that the collation condition is satisfied. In this manner, one can define $\bar{u}$ over all sets $W_i$
satisfying the indifference condition. Let $W^1$ denote the union of all such sets.

Note that $W^1$ cannot have a non-empty intersection with any $W_j$ where preferences are comparable by $\prec$ (there are $p_j, q_j \in P(W_j)$ with $p_j \prec q_j$).
So consider separately the collection $\{W_k\}$ where each $p_k \prec q_k$. Define the real-valued functions $a_k$
as $a_k(x)=1$ if $x\in W_k$ and $a_k(x)=0$ otherwise. Similarly for $W^1$ above define $b(x) = 1$ if $x\in W^1$ and $b(x) = 0$ otherwise.
Next, define $a_{ij}(x) = 1$ if $x\in W_i\cap W_j$ and $a_{ij}(x) = 0$ otherwise.
Each $a_k$ is continuous because for any open $V\subseteq [0,1]$ $a_k^{-1}(V)$ is open -- being equal to either $X$, $\emptyset$, $W_k$,
or its complement $W_k^c$ (by Assumption~5, $W_k^c$ is also open). By a similar argument, $b$ and $a_{ij}$ are continuous as well.
Now define the following variable lotteries:
$$\pi(x) = \sum_k a_k(x)q_k(x)- \sum_k\sum_{j>k} a_{kj}q_j+b(x)\delta_1(x),$$
$$\theta(x) = \sum_k a_k(x)p_k(x)- \sum_k\sum_{j>k} a_{kj}p_j+b(x)\delta_1(x).$$
Both $\pi$ and $\theta$ are continuous as well, and so in $P(X)$.

Let $W^2 = X-W^1$. Then, $W^2 \Vdash \theta|_{W^2} \prec \pi|_{W^2}$ and, from the argument of Corollary~1, there must exist a representing function
$u$ over $W^2$. Noting that $W^1 \cap W^2 = \emptyset$, $u$ can be extended to $W^1 \cup W^2$ by combining with the
function defined over $W^1$ above.

Necessity of Assumptions 1-3 is proved next. If there is a function $u$ which represents preferences in the sense of equation (3), then
Assumptions 1-3 must be satisfied.
Since $u(p)$ and $u(q)$ are continuous real-valued functions, if $W \Vdash u(p) < u(q)$
then there is no open subset of $W$ for which $u(q) < u(p)$, and asymmetry must hold.
For negative transitivity, if $W \Vdash u(p) < u(q)$, and given $r$ with expected utility $u(r)$, there is an open cover
$\{W_i\}$ of $W$, such that for each $i$, $W_i \Vdash u(r) < u(q)$ or $W_i \Vdash u(p) < u(r)$.
The restriction operation is order-preserving in the case of the ordering by $<$ of continuous real-valued functions. So, ordering of variable lotteries by $\prec$
must be preserved upon restriction as well. The collation condition also follows from the same property of real numbers (see discussion after Assumption~1).
For independence, if $W \Vdash u(p) < u(q)$ and $r\in P(W)$ with $u(r)$, then
$W \Vdash u (ap + (1-a)r) < u(aq + (1-a)r)$ for continuous $a:W\rightarrow \mathbb{R}$, and so
$W \Vdash ap+(1-a)r \prec aq+(1-a)r$. For continuity, consider $W \Vdash u(p) < u(q) < u(r)$.
To produce a continuous $a$ such that $u(q) < au(r)+(1-a)u(p)$, take $a:W\rightarrow \mathbb{R}$ defined by
$a = (u({1 \over 2}q + {1 \over 2}r) - u(p))/(u(r)-u(p))$. The second part is similar.
\end{proof}

The necessity part of the proof above holds in the case of Theorem~1 as well. It is however unclear if Assumption~4 is implied by the representation
with real-valued functions. However, in the presence of classical logic, minimal comparability must hold because
$ (\forall p,q \in P(X), p\sim q) \vee (\exists p,q \in P(X), p\prec q) $ is always true by the law of the excluded middle.

\begin{proof}[Proof of Corollary 2]
The condition that $\delta_1 \prec \delta_n$ over $X$ implies Assumptions 4 and 7. So, by Corollary~1, there is a mapping
$u(\cdot)$ that represents preferences over $X$.
Let $u(\delta_n)$ be the constant function with value 1 over $X$, and let $u(\delta_1)$ have value 0.
For all constant functions $\alpha: X \rightarrow [0,1]$, from Lemma~4,
\[ u(\alpha \delta_n + (1-\alpha)\delta_1) = \alpha u(\delta_n) + (1-\alpha)u(\delta_1) = \alpha.\]
Now consider some constant distribution $\delta_i$ with $\delta_1 \precsim \delta_i \precsim \delta_n.$
From Lemma~2, there is a unique $\beta:X \rightarrow [0,1]$ such that
\[ \delta_i \sim \beta \delta_n + (1-\beta) \delta_1.\]
Consequently, from Lemma~4, $u(\delta_i) = \beta$. Observe that $\delta_i$ and $\alpha \delta_n + (1-\alpha)\delta_1$
are both constant distributions (for $\alpha$ a constant function). Consequently, Assumption~7 implies that
one of the following must hold over all of $X$ for every constant function $\alpha$: $\delta_i \prec \alpha \delta_n + (1-\alpha)\delta_1$,
$\alpha \delta_n + (1-\alpha)\delta_1\prec \delta_i$ or $\alpha \delta_n + (1-\alpha)\delta_1\sim \delta_i$. For every constant $\alpha: X \rightarrow [0,1]$,
 $\forall x\in X, \beta(x)<\alpha(x)$, or, $\forall x\in X, \beta(x)=\alpha(x)$, or $\forall x\in X, \beta(x)>\alpha(x)$.
In other words, if $\beta$ intersects any $\alpha$, it must coincide with it. So $u(\delta_i)=\beta$ is a constant function.
The result extends to any constant distribution upon noting that $u(p) = \sum_z u(\delta_z)p_z$ (from Lemma 5).
\end{proof}

\newpage
\appendix

\section*{Appendix}
\renewcommand{\thefigure}{A\arabic{figure}}
\setcounter{figure}{0}
\pagenumbering{arabic}
\renewcommand*{\thepage}{A\arabic{page}}

\subsection*{A Brief Survey of Constructive Mathematics and Sheaf Theory}
The purpose of this appendix is to make this paper relatively self-contained. To this end, I provide an introduction to relevant ideas in constructive mathematics and sheaf theory. I also elaborate here on the link between topology and computation, which plays a central role in this paper. The tone will be informal, with the focus being on providing the intuition behind key results. More formal and complete surveys of the subject can be found in the various references cited in this section. \citet{bauer2017} is a good short introduction, while \citet{goldblatt1982}, \citet{bridgesrichman1987}, \citet{maclane1996}, and \citet{vickers1996} provide more comprehensive surveys. The sheaf formalism is used here as a device for modeling preferences that are not fully formed. Rather, they are the outcome of some process of computation, observation, or measurement. The price to pay for use of this formalism is that arguments need to be constructive.  So, I will begin by describing what constructive mathematics is. This will be followed by a comparison of classical and intuitionistic logic (the latter being the logic used in constructive mathematics). Following this, I present some examples of reasoning about phenomena for which the natural logical system is intuitionistic. I then discuss sheaves and their properties, emphasizing their nature as a universe of variable sets. Finally, I show how mathematical statements are made within this universe, and how the truth of statements is to be deduced.

\subsubsection*{Constructive Mathematics and Intuitionistic Logic}
Constructive mathematics is done without the {\it law of the excluded middle\/}: this principle of logic states that for every proposition
$p$, either $p$ is true or its negation $\neg p$ is true (more formally, $\forall p, p\vee \neg p$). A consequence of the law is that proof by contradiction is unacceptable. Such a proof rests on the notion that if a proposition can be shown to be not false, then one can conclude it is true (i.e., $\neg \neg p \Rightarrow p$). Also unacceptable is the axiom of choice, since this has been shown to  imply the law of the excluded middle \citep[see][ p.~483]{bauer2017}.
The name ``constructive mathematics" derives from the fact that objects proved to exist by constructive methods
are {\it algorithmically realizable\/} or {\it computable\/}. In other words, if something is asserted to be true, there is an algorithm witnessing its truth; and if something is asserted to exist, it is possible to construct it through an explicit computation. A non-constructive proof can sometimes be replaced by a constructive proof. However, there are instances where there can be no algorithm to construct an object purported to exist. In such instances, the culprit has been found to be the law of the excluded middle.

An oft-cited example of a proposition that can be asserted in classical mathematics, but not in constructive mathematics, is the assertion that every real number is either equal to zero or different from zero: $(\forall x \in \mathbb{R}) (x = 0 \vee x \neq 0)$. This classically valid assertion -- true by the law of the excluded middle -- is constructively suspect. There can be no general method for determining, for an arbitrary real number $x$, if $x=0$ or $x \neq 0$. To see why, consider the still unsolved Goldbach conjecture, which states that every even natural number greater than $2$ is the sum of two prime numbers \citep[see][p.~98]{bridges1999}.
Consider the ordered sequence of even natural numbers greater than $2$ ($\{4, 6, 8, \ldots \}$) and construct
a new sequence $\{\alpha_k\}$ as follows. For each member of the former sequence it is possible to determine finitely whether it is the sum of two prime numbers or not. If the $k$-th member of the sequence is the sum of two primes, set $\alpha_k =0$; if not, set $\alpha_k =2^{-k}$. Define $\alpha = \sum_{k=1}^\infty \alpha_k$. So we have constructed a real number $\alpha$ such that the Goldbach conjecture is true if and only if $\alpha =0$. If the conjecture fails,
then $\alpha > 0$. Note that $\alpha$ is a real number because $\sum_{k=1}^n \alpha_k$ provides us with a convergent sequence of rational numbers,
yielding as close an approximation to $\alpha$ as we want (for $n$ large enough). Any general method for determining whether an
arbitrary real number is equal to zero or different from zero could be used to resolve the Goldbach conjecture.\footnote{There are other versions of this
result. For instance, \citet{turing1936}, p.~248, states a similar result about his ``computable'' real numbers.}
It is worth noting that the Goldbach conjecture has been confirmed for integers less that $4 \times 10^{18}$.\footnote{See the Wikipedia entry on Goldbach's conjecture.} By checking specific instances one can never hope to affirm the conjecture,
however it is possible to refute it by finding a specific instance of failure. This illustrates the important point that
the observational content of a proposition and that of its negation can be quite different.
Examples such as the above, which show that solutions for certain problems (such as whether $x$ is, or is not, zero) would imply solutions for other difficult problems (such as the Goldbach conjecture) for which no solution is available, go under the
name of ``Brouwerian counterexamples."

Constructivists such as Brouwer questioned the meaning of existence in circumstances where the object claimed to exists is not algorithmically realizable. They deny the claim that to assert existence of an object it is sufficient to show that the assumption that the object does not exist creates a contradiction. A number of familiar results in mathematics (the intermediate value theorem, the extreme value theorem, etc.) turn out not to be constructively valid in this sense. Take the example of the extreme value theorem. Here, one can use a counterexample closely-related to the previous one: one cannot in general decide, for
arbitrary real number $x$, whether $x\geq 0$ or $x\leq 0$ \citep[see][p.~100]{bridges1999}.
Assume that $f$ is a continuous function on $[0,1]$. The function is assumed to be
computable in the sense that one can obtain arbitrarily close approximate to $f(x)$ by taking suitably close approximations of the argument $x$. It turns out that one cannot always approximate a point $x^*$ for which the function attains its maximum.The reason is related to the fact that the {\em argmax\/} of $f$ can be discontinuous, and around the points of discontinuity the potential maximum points can be far apart. Consider for instance $f$ defined on $[0, 1]$ by $f(x) = (1+\alpha) x + (1-\alpha) (1-x)$.
Here, $\alpha$ is a real-valued parameter which can be approximated to an arbitrary degree of precision.
As a result, approximations to $x$ can be translated into those of the value of $f$ at $x$.
Now, if $\alpha > 0$, then $x^*=1$. If $\alpha <0$, then $x^*=0$.  If $\alpha = 0$, then any $x^*\in [0,1]$ is a maximum point.
An algorithm to compute approximations of a maximum point $x^*$ would enable us to determine for arbitrary $\alpha$ if
$\alpha \geq 0$ (when $x^*$ is in a neighborhood of 1 that excludes 0) or if $\alpha \leq 0$ (when $x^*$ is in a neighborhood of 0 that excludes 1). If this were true for arbitrary $\alpha$, we would have a general method for determining for an arbitrary real number $x$, whether $x\geq 0$ or $x \leq 0$. .
In contrast to the maximum point, the maximum {\it value\/} of $f$ can be computed (e.g., by taking the maximum value
over a suitably fine grid of points of $[0,1]$). A similar phenomenon holds for the intermediate value theorem -- it is not possible to
compute the zero of a continuous function $f:[0,1] \rightarrow \mathbb{R}$ with $f(0)<0$ and $f(1)>0$ if the value of the function hovers around 0 for a range of values of $x$ (e.g., because $f$ has a flat stretch). However, the situation isn't hopeless. In both cases, approximations {\it are\/} computable. It is possible to find an approximate maximum point $x^\epsilon$ for which
the value of the function, $f(x^\epsilon)$, is within some bound $\epsilon$ of the maximum value (respectively, zero value) of the function.
It is also possible to restrict the class of functions to obtain positive results. In the case of the intermediate value theorem, if $f$ is locally non-constant, its zero becomes computable. In his landmark treatise, \citet{bishop1967} shows that much of classical analysis can still be done while adhering to constructive principles. In particular, constructive mathematics does not involve having to give up the most valuable parts of classical analysis, as Hilbert had feared.\footnote{``Taking the principle of excluded middle from the mathematician
would be the same, say, as proscribing the telescope to the astronomer
or to the boxer the use of his fists,'' \citep[see][]{bauer2017}.}

I mentioned above that constructive proofs forbid use of the law of the excluded middle. The logic underlying constructive proofs goes under the name of {\it intuitionistic logic\/}. For the logic of propositions, in classical logic there is the {\it propositional calculus\/} \citep[see, for instance, ][]{goldblatt1982, maclane1996}. In this latter system, propositions $p, q, \ldots $ are combined under operations “and”, “or”,  “implies”, and “not” (represented by their corresponding symbols $\wedge, \vee, \Rightarrow,$ and $\neg$) to construct {\it sentences\/}. Propositions $p, q, \ldots $ are themselves sentences, and new sentences can be created using the logical connectives following certain rules: (1)~if $\alpha$ is a sentence, then so is $\neg \alpha$, (2) if $\alpha$ and $\beta$ are sentences, then so are $\alpha \vee \beta$, $\alpha \wedge \beta$, and $\alpha \Rightarrow \beta$. To assign meaning to the sentences, we may think of propositions $p, q, r, \ldots$ as being represented by sets $P, Q, R \ldots$ as their {\it truth values\/},  which are subsets of some set $X$. Think of $p$ being true when $x\in P$. The propositional connectives $\wedge$ and $\vee$ become intersection ($\cap$) and union ($\cup$). In other words, $p\wedge q$ is true if $x\in P\cap Q$, and $p\vee q$ is true if $x\in P\cup Q$. Corresponding to  negation, the set complement is the truth value (i.e. $P^c$ for $\neg p$) with $\neg p$ true if $x\in P^c$.  For implication, $p\Rightarrow q$, the truth value is $(Q \cup P^c)$. In this way, the collection of all subsets of $X$ constitute the set of possible {\it truth values\/} of the sentences, with the understanding that $X$ signifies {\bf True\/} and $\emptyset$ signifies {\bf False\/}. The truth values specify when or where in $X$ a proposition is true, with $X$ meaning always or everywhere, and $\emptyset$ meaning never. The propositions and connectives are augmented by certain axioms ( \citep[e.g.,][p.~131]{goldblatt1982}). Among these is the law of the excluded middle -- for any sentence $\alpha$, $\alpha \vee \neg \alpha$).\footnote{In the alternate version, with truth values as subsets of $X$, the law of the excluded middle can be stated as follows: if $A$ is the truth value of $\alpha$ and $A^c$ of $\neg \alpha$ then $A\cup A^c = X$. In other words, one of $\alpha$ or $\neg \alpha$ is always true.} Additionally, there is a rule of inference called
{\it detachment\/} or {\it modus ponens\/} -- from $\alpha \Rightarrow \beta$ and $\alpha$ derive $\beta$. Derived sentences are the theorems of the system. The set $X$ and its subsets (together with the above operations) is a Boolean algebra, and a model for the propositional calculus.

Intuitionistic propositional calculus (or Heyting algebra) differs from the propositional calculus in not having the law of the excluded middle as an axiom \citep{goldblatt1982, maclane1996}. It is a different algebraic system, although with the same operations. Importantly, the typical model for this logical system is now not the set of all subsets of $X$, but the set of all {\it open\/} subsets of some topological space $(X, \mathcal{O}(X))$. Now the open sets in $\mathcal{O}(X)$
are the potential truth values of the propositions within the system. In the constructivist spirit, truth now corresponds to the verifiability of a proposition. Open sets correspond to the confirmation of a proposition (alternatively, the proposition is semi-decidable or observable). The relationship between open sets and finitely observable properties has been developed at length by \citet{vickers1996}, who formulates the closely related {\it propositional geometric logic\/} as the basis for the logic of finite observations.
Starting with open set truth values of propositions, it is possible to obtain truth values of all sentences that can be constructed using the propositions and logical connectives, but there are important differences with classical logic (and the algebra of all subsets of $X$). Intersection and union are as before, but implication $p \Rightarrow q$ has as its truth value the {\it interior\/} of
$(Q \cup P^c)$ (denoted Int $(Q \cup P^c)$) and negation $\neg p$ has truth value Int($P^c$). Clearly, any logic of finite observation
must exclude infinite conjunctions, which the definition of a topology does (only the intersection of finite collections of open sets are
required to be open). Infinite disjunctions are however allowed.

To appreciate the significance of this change with the move from classical logic to intuitionistic logic -- i.e., the move from all subsets of a set $X$ to the open subsets of $X$ -- consider again
the example of determining if an arbitrary real number $x$ is equal to $0$ or different from zero. If $x$ is different from $0$, this can be verified in finite time---we just keep generating better and better approximations to the number.
If the number is indeed different from $0$, this will eventually be revealed. However, consider the case of $x=0$. This cannot, in general, be finitely confirmed. This is an example of a proposition which can be
finitely confirmed ($x \neq 0$), but its negation ($x=0$) cannot be. The truth value of $x \neq 0$
is $(-\infty, 0)\cup (0,+\infty)$, but the truth value of $x=0$ is Int($\{0\}$) $ = \emptyset$. With truth as verification, the fact that a proposition can be finitely confirmed does not mean its negation can also be finitely confirmed.
One can never confirm $x=0$. A similar issue arises for classical implication (indeed, observe that the truth value of $\neg p$ and $p \Rightarrow$ {\bf False\/} are the same). If the set of truth values is
$\{X, \emptyset\}$ then for any $U \in \{X, \emptyset\}$both $U$ and $U^c$ are open, and $U \cup U^c = X$. However, in general excluded middle could fail:
typically, for open $P$,  $P \cup \mbox{Int}(P^c)$ need not equal all of $X$. The classical case now appears as a special case of intuitionistic logic where the set of truth values have the property that for all $U \in \mathcal{O}(X)$, $U \cup U^c = X$
(an example being the discrete topology). This last condition -- $\forall U \in \mathcal{O}(X), (U \cup U^c = X)$ -- is an additional
requirement of classical logic. Constructive mathematics involves “making do with fewer axioms” and, as a result,
constructively valid theorems are always valid in classical mathematics, but the reverse need not be true.

The above is a highly abbreviated account of classical and intuitionistic logic, but the main objective here is to highlight the contrast between truth values in the two systems -- i.e., between all subsets of $X$ and all {\it open\/} subsets of $X$. \citet{vickers1996} presents a number of examples of reasoning about phenomena where an intuitionistic logic applies.
\begin{enumerate}
\item One example is the measurement of a physical quantity, such as someone's height. Due to physical constraints, it may only be possible to identify an open interval within which the height lies (e.g., $(180\mbox{cm} - \varepsilon,
180\mbox{cm} + \varepsilon)$). With improvements in the accuracy of measurement devices, one could make $\varepsilon$ smaller, but it may not be physically possible to bring it down to zero.
One cannot then assert the height to be exactly $180\mbox{cm}$, but other assertions about height (e.g., ``height is greater than $170\mbox{cm}$'') are possible. A reasonable choice for the set of possible assertions about
the value of height is a collection of open intervals.
\item A second example is that of generalization from empirical observation (as in scientific hypotheses).
The statement ``All ravens are black'' can be shown to be false (by producing a non-black raven), but can never be asserted to be definitely true based on finite observation alone.
In this case, the law of the excluded middle clearly fails if verification is the condition for truth.
\item A third example -- central to the purposes of this paper -- if of objects undergoing change. As a variation on the height example above, instruments of measurement might limit our ability to assert {\it exactly when\/}
a growing child reached a height of $100\mbox{cm}$. Here, truth values equal the points in time when assertions about the child's height were verifiably true. Open intervals of the real line would appear to be the appropriate choice.
\end{enumerate}

A more profound example of change is the accumulation of knowledge over time.
I consider mathematical knowledge, but the idea is completely general. In the intuitionistic framework mathematical assertions are a record of constructions made at specific points in time.
Consider a currently unresolved mathematical problem -- the {P} vs. {NP} problem.\footnote{This is one of the Millennium prize problems identified by the Clay Math Institute, with a solution to be awarded a purse of \$1 million. If P $=$ NP, the implication would be that a class of problems, such as the Traveling Salesman Problem, whose solutions
are ``easy to check" (NP) also have solutions that are ``easy to find" (P). The problem is to determine whether P $=$ NP or P $\neq$ NP.}
At this point in time it is unknown whether {P} $=$ {NP} or {P} $\neq$ {NP}. So neither of these could be asserted right now. In the intuitionistic setting, one also cannot assert ({P} $=$ {NP} $\vee$ {P} $\neq$ {NP}). There is a possible future in which {P} $=$ {NP} is asserted, and a different possible future where {P} $ \neq$ {NP} is asserted. But there is also the possibility that the problem remains forever unresolved. The situation is summarized in Figure~A1. In the left panel, (a), stage~1 is defined by all known mathematical assertions plus {P} $\neq$ {NP}, stage~2 is defined by all known mathematical assertions plus {P} $=$ {NP}, and at stage~0 no assertion is made about the {P} vs. {NP} problem. There is a partial ordering of the stages of knowledge.  Stages $1$ and $2$ are {\it more refined\/} than $0$, in the sense that they include all the mathematical assertions that 0 does, and more. Stages $1$ and $2$ are not similarly comparable by this ordering. This can be written as follows: $0 \sqsubseteq 1$ and $0 \sqsubseteq 2$. Additionally, for $x \in \{0, 1, 2\}$, $x \sqsubseteq x$.  A second example is presented in the right panel (b). Here there are only two stages of knowledge with $0 \sqsubseteq 1$, $0 \sqsubseteq 0$, and $1 \sqsubseteq 1$ (e.g., stage $1$ is defined by the assertion ``the P vs. NP problem has been resolved").
In this way, one could have arbitrarily many stages of knowledge ordered by refinement. An important requirement
for the stages of knowledge depicted in Figure~A1 is that an assertion once made can never become false.
This property is known under the name {\it persistence of truth in time\/}.
So, $0$ is defined not by the assertion that ``the {P} vs. {NP} problem is unsolved,'' which
would become false if the problem was ever solved,
but by the absence of any assertion about the problem. Stage $0$ involves perpetually being in a state where one
cannot rule out that $\{1\}$ or $\{2\}$ might be confirmed. One can only make those assertions in $0$ that continue to be true in both $1$ and $2$ (i.e., are true for $\{0, 1, 2\}$).

Ordering by refinement gives rise to a topology by taking the upper sets to be open
(i.e., if $x$ is included in an open set $U$, and $x \sqsubseteq y$,
then $y$ is also included in $U$).\footnote{This is called the Alexandrov topology.}
Corresponding to the left panel (a) the set of stages is $X = \{0, 1, 2\}$ and the collection of open sets is
$\mathcal{O}(X) = \{ \{0,1,2\}, \{1\}, \{2\}, \{1,2\}, \emptyset\}$. Similarly, corresponding to (b),
$X = \{0, 1\}$ and $\mathcal{O}(X) = \{ \{0,1\}, \{1\}, \emptyset\}$. The latter is known as the Sierpi\'nski space.
Open sets may now be thought of as observable properties that can be asserted. For (b), in state $0$, one can only assert $\{0,1\}$, whereas in stage $1$, one can assert both $\{0,1\}$ and $\{1\}$. Having asserted only $\{0,1\}$ it is not possible to rule out that $\{1\}$ will eventually be observed (note that Int($\{1\}^c$) $= \emptyset$, so one can never assert that $1$ does {\it not\/} occur).
Similarly, the topology in (a), has $0$ satisfying $\{0,1,2\}$, $1$ satisfying $\{0,1,2\}\wedge \{1,2\} \wedge \{1\}$ and
$2$ satisfying $\{0,1,2\}\wedge \{1,2\} \wedge \{2\}$.\footnote{Indeed, it is possible to begin with open sets as the basic primitive,
and points as the derived construct. The points of $\mathcal{O}(X)$ then correspond to the elements generating prime principal ideals.
To repeat a slogan from \citet{vickers1996} the open sets are the observations, and a point is what is being observed.}
An important observation is that these topologies do not satisfy the Hausdorff separation axiom, which is typical of such examples.

\begin{figure}
\begin{center}
\begin{tikzpicture}[line cap=round,line join=round,>=triangle 45,x=1cm,y=1cm, scale=0.75]
\draw [line width=1pt] (10,4)-- (10,1);
\draw [line width=1pt] (4,1)--(2,4);
\draw [line width=1pt] (4,1)--(6,4);
\node [color=black] at (10,0.5) {$0$};
\node [color=black] at (10,-0.5) {(b)};
  \node [color=black] at (10,4.5) {$1$};
  \node [color=black] at (4,-0.5) {(a)};
  \node [color=black] at (4,0.5) {$0$};
  \node [color=black] at (2,4.5) {$1$};
  \node [color=black] at (6,4.5) {$2$};
\end{tikzpicture}
\caption{Stages of knowledge}
\end{center}
\end{figure}

Now consider functions defined on $X = \{0, 1\}$. Suppose $f: X \rightarrow \{0,1\}$ with $f(0)=f(1)=0$ (and let
$\mathcal{O}(X) = \{ \{0,1\}, \{1\}, \emptyset\}$). Assume the codomain has the
discrete topology. One may ask ``When is $f$ equal to 0?'' More formally, what is the truth value of $f=0$?
With the notation $\llbracket \phi \rrbracket$ for the truth value of $\phi$, one has  $\llbracket f=0\rrbracket = \{0, 1\}$. Now consider
the function $g(0)=1$ and $g(1)=0$. In this case, $\llbracket g=0\rrbracket = \{1\}$, which is open.
How about $\llbracket g=1\rrbracket$?
If it were possible to confirm $g=1$ one would have confirmation that the
stage is $0$. But the choice of topology rules this out, because $\{0\}$ is not open (the confirmation $g=1$ violates
the observability restrictions encoded in the topology). Hence, it is never possible to confirm that $g=1$ and
$\llbracket g=1\rrbracket = \emptyset$. So,
$\llbracket g=0 \vee g\neq 0 \rrbracket = \{1\}$ and excluded middle fails. In general, for any sentence about
the value of a function, its truth value is the largest open set for which the sentence holds.
In this spirit, for observational content to be preserved by a map $f$ between topological spaces $X$ and $Y$, it must be the case that every open $V \subseteq Y$ satisfy the requirement that $f^{-1}(V) = U$ is an open set. This is just the definition of continuity of $f$. In the examples above $f$ is continuous, whereas $g$ is not.

\subsubsection*{Sheaf Theory}
The rest of the appendix deals with the topic of sheaves. The approach taken in this paper is to have lotteries vary continuously over some space $X$. In other words, they are functions from $X$ to a space of lotteries. The collection of open sets in $X$, $\mathcal{O}(X)$, serve as the truth values specifying {\it where\/} in $X$ some property of these variable lotteries is satisfied. For instance, if $f$ and $g$ are two such functions, it may be that the decision-maker asserts preference for $f$ over $g$
($g \prec f$, in the usual notation) upon the assertion of some open set $U \subseteq X$ (e.g., because $g(x) \prec f(x)$ for every $x\in U$).
I assume that for open $V \subseteq U$ $g \prec f$ also holds when $V$ is asserted (this is, for the examples in Figure~A1, just the property of persistence of truth in time).\footnote{The paper discusses cases where such an assumption is not reasonable.} Such a property holds for continuous real-valued functions $f$ and $g$ with ordering defined by $<$. If $f(x)<g(x)$ for all $x\in U$, and $V \subseteq U$, then $f(x)<g(x)$ for all $x\in V$. This and other properties needed for the purposes of this paper are captured in the definition of a {\it sheaf\/} on a topological space $X$. Below, I repeat some standard definitions from \citet[see][p. 488]{bauer2017} and \citet[][Ch.~II]{maclane1996}.

A sheaf of sets, $F$, is a set $F(U)$ indexed by $U\in \mathcal{O}(X)$, and satisfying certain properties to be enumerated below. The
collection $\mathcal{O}(X)$ is a poset, ordered by inclusion ($\mathcal{O}(X)$ can also be viewed as a category in which case there is an arrow from $V$ to $U$ when $V \subseteq U$ (written as $V \hookrightarrow U$ and called inclusion). Corresponding to each inclusion arrow $V \hookrightarrow U$ is an arrow from $F(U)$ to $F(V)$,
an operation of restriction $\rho_{UV}$ which send $x\in F(U)$ to its restriction $x|_V \in F(V)$. Note the reversal in the direction of arrows from $V \hookrightarrow U$ to $F(U) \rightarrow F(V)$.\footnote{Formally, letting $\mathcal{O}_X$ denote $\mathcal{O}(X)$, $F$ is a functor $F: \mathcal{O}_X^{op}\rightarrow {\bf Set}$. Here $\mathcal{O}_X^{op}$ is the category with objects as in ${\mathcal{O}_X}$, but with arrows reversed. $F$ is a sheaf when it satisfies the additional conditions listed below.} An open cover, or covering, of $U\in \mathcal{O}(X)$ is a collection of open sets $\{U_i\}_{i\in I}$ such that $\bigcup_i U_i = U$. The two defining conditions of a sheaf can now be stated:
\begin{enumerate}
\item We have $\rho_{UU}=${\bf id\/} and,  for $W \subset V \subset U$, $\rho_{VW}\circ \rho_{UV}  = \rho_{UW}$ (i.e., for $x\in F(U)$, $x|_U = x$, and $(x|_V)|_W = x|_W$).
\item If $\{U_i\}_{i\in I}$ is an open cover of $U \subseteq X$ and $\{x_i| x_i\in F(U_i) \}$ is a collection of elements satisfying $x_i|_{U_i\cap U_j} =
x_j|_{U_i\cap U_j}$ (i.e., they agree on overlaps) then there is a unique $x\in F(U)$ such that $x|_{U_i} = x_i$ for all $i\in I$.
\end{enumerate}
The first is a transitivity property of restriction. The second property is called the {\it gluing property.\/} In this paper, elements of the set $F(U)$ will always be continuous
functions defined over $U$. Additionally, in section~2 of the paper, I will be defining an ordering $\prec$ over elements of $F(U)$.

An example of a sheaf is the set of all continuous real-valued functions defined over $U\in \mathcal{O}(X)$:
\[ C(U) = \{ f: U \rightarrow \mathbb{R} | f \ \mbox{is continuous.} \} \]
It is clear that the restriction property is satisfied since, for $f:U \rightarrow \mathbb{R}$ and open $V \subset U$, $f|_V$ is continuous.
The gluing property holds as well---given an open cover $(U_i)_{i\in I}$, if the functions $f_i:U_i \rightarrow \mathbb{R}$ are continuous for
all $i\in I$, then there is at most one continuous function $f:U\rightarrow \mathbb{R}$ with $f|_{U_i}=f_i$ which exists if and only if the $f_i$
match on overlaps ($f_i(x) = f_j(x)$ for $x\in U_i\cap U_j$ for all $i,j\in I$). In the terminology of \citet{maclane1996}, continuous
functions can be uniquely collated. In contrast, the set $K(U)$ of all constant real-valued functions with domain $U$ is {\em not\/} a sheaf, because
it fails to satisfy the gluing condition.\footnote{Let $X$ be the union of two disjoint sets $U_1$ and $U_2$. Suppose $K(U_1)$ and $K(U_2)$ are sets of
constant functions. Two functions $f_1 \in K(U_1)$ and $f_2 \in K(U_2)$ with $f_1(x)\neq f_2(x)$ satisfy the gluing condition, but there is no element of
$K(U_1\cup U_2)$ of which $f_1$ and $f_2$ are restrictions.} On the other hand, if $K(U)$ is defined to be the set of all {\em locally\/} constant functions (i.e.,
for every $f\in K(U)$ and $x\in U$ there is an open neighborhood $U_x$ of $x$ over which $f$ is constant) then $K$ is a sheaf.

A mapping $\eta$ between sheaves $F$ and $G$ is called a {\it natural transformation\/}.
It is a collection of maps $\eta_U: F(U)\rightarrow G(U)$ for each $U\in \mathcal{O}(X)$
and restriction maps for $F$ ($\rho_{UV}$) and $G$ ($\xi_{UV}$) with the property that
$\xi_{UV} \circ  \eta_U =\eta_V \circ \rho_{UV}$ (i.e., for any $x \in F(U)$, $\eta_U(x)|_V = \eta_V(x|_V)$).
Sheaves on a topological space $X$ and their natural transformations form a {\it category\/}, denoted Sh($X$).
Additionally, this category satisfies certain conditions
that make it a {\it topos\/}.\footnote{A general discussion of categories and toposes is omitted because these are not required in other sections of the paper.
The properties of Sh($X$) are all that is needed. \citet{maclane1996} is a good source for both topics.}
Roughly, Sh($X$) is a generalization of the familiar universe of sets within which everyday mathematics is done (this latter category is denoted by {\bf Set\/}).
Sh($X$) is obtained by having
objects within the category of sets vary continuously over the topological space $X$.\footnote{Lawvere and Tierney made the important discovery
that a topos is a mathematical universe within which one can carry out familiar set theoretic constructions. They showed that the universe of sets happens to be a topos, but
other mathematical worlds could also be constructed as toposes. Among these is Sh($X$), the world of variable sets.} If we take a singleton space $1= \{*\}$
with the open sets $\{\{*\},\emptyset\}$ as its truth values (corresponding to {\bf True} and {\bf False} respectively), then Sh($1$) becomes the category of sets
(or {\it constant\/} sets, to distinguish this from sheaves where sets undergo variation).
The important point, for present purposes, is that ``the logic governing truth in a topos is \ldots intuitionistic logic'' and ``a topos can be viewed as an
intuitionistic universe of sets" \citep[][p. 268]{maclane1996}. This is true for the topos Sh($X$), where the open sets $\mathcal{O}(X)$ constitute the
truth values. So, for $f, g \in C(X)$, a statement like $f<g$ will have truth value $\llbracket f<g \rrbracket$, the largest open set in $\mathcal{O}(X)$
over which $f<g$. Note that, over $\llbracket f<g \rrbracket$, $f<g$ is {\em verifiably\/} true. Upon restriction to $U\subset X$ the truth values are
all $V\in \mathcal{O}(X)$ with $V \subseteq U$, which we may denote as $\mathcal{O}(U)$.

Truth values of more complex statements are defined by extending the ideas above. The distinction between classical and
intuitionistic {\it propositional\/} logic was discussed above. {\it First order logic\/} extends classical propositional calculus to also include variables,
predicates and quantifiers. In a similar manner, first order intuitionistic predicate calculus extends intuitionistic propositional calculus while continuing
to conform with constructive principles.
In both the classical and constructive setting, there are standard rules for constructing sentences or formulae. For instance, the following is an
example from {\bf Set\/}:
$(\forall f:[0,1]\rightarrow [0,1])[f \mbox{continuous} \Rightarrow (\exists x^*\in [0,1])f(x^*)=x^*]$
(of course, ``$f$ continuous'' is just shorthand for the familiar formal statement defining continuity). Such sentences (involving constants, variables, functions,
and quantifiers), can also be
constructed in Sh($X$) where they refer to entities that exist within Sh($X$) (i.e., objects like those in {\bf Set\/}, but now varying over $X$).
Rules for constructing formulae or sentences in a general topos are provided by the Mitchell--Benabou language, \citet[][Ch.~VI]{maclane1996}).
In Sh($X$), one needs to be explicit about the domain of definition of an element $f$ of a sheaf $F$ (i.e., the set $U \in \mathcal{O}(X)$ over which $f$ is
defined), as well as the object in Sh($X$) that $f$ is an element of (i.e., $F$). When it comes to determining the truth of statements, the rules of inference in
Sh($X$) are typically the rules for the first-order intuitionistic predicate calculus. Classical logic only applies in special cases (e.g., in Sh($1$),
where the set of truth values is a Boolean algebra). A consequence is that
theorems with constructive proofs can be carried over to all sheaves, whereas classical results carry over only to special cases.

Many constructions from {\bf Set\/} can be carried out in Sh($X$) as well. An example is the Dedekind construction of the real numbers.
Within set theory, one way in which real numbers are constructed is via Dedekind cuts.
Letting $\mathbb{Q}$ denote the rational numbers, a Dedekind cut comprises of two sets of rational numbers $(L,U)$ which satisfy the following properties:
\begin{enumerate}
\item {\bf Inhabited.\/} $\exists v\in \mathbb{Q} (v \in L) \wedge \exists w\in \mathbb{Q} (w \in U).$
\item {\bf Unbounded.\/} $\forall v, w \in \mathbb{Q} (v < w \wedge w \in L \Rightarrow v \in L) \wedge (v < w \wedge v \in U \Rightarrow w \in U).$
\item {\bf Open.\/} $\forall v \in \mathbb{Q} (v\in L \Rightarrow (\exists w\in \mathbb{Q})(w>v \wedge w\in L)$ and \\  $\forall v \in \mathbb{Q} (v\in U \Rightarrow (\exists w\in \mathbb{Q})(w<v \wedge w\in U).$
\item {\bf Arbitrarily close.\/} $\forall v, w \in \mathbb{Q} (v<w \Rightarrow v\in L \vee w\in U).$
\item {\bf Disjoint.\/} $ \forall v \in \mathbb{Q} \neg (v\in L \wedge v \in U).$
\end{enumerate}
Starting with the natural number object in Sh($X$), which are locally constant functions with natural number values,
one can construct a sheaf of rational numbers there. Then the Dedekind reals can be defined by a suitable formula in the
Mitchell--Benabou language that is essentially a conjunction of the five
statements about $(L,U)$ above. \citet{maclane1996} in Theorem 2 (p.~322) show that the object of real numbers so constructed
is isomorphic to the sheaf $C$ of continuous
real-valued functions on the space $X$. In other words, it is the set of (classical) real numbers varying continuously over $X$.
There is an alternative way of constructing the reals as limits of Cauchy sequences of rational numbers (with two sequences
considered equivalent if their difference converges to zero).  In set theory, the two approaches lead to the same real number object.
However, this is not generally true for Sh($X$), where the Cauchy reals are the locally constant real-valued functions.

Corresponding to the formal language for Sh($X$) there is the {\em sheaf semantics\/} which tells us when a formula constructed in Sh($X$) is true.
The semantics is formulated in terms of a ``forcing relation'', $\Vdash$. Its meaning and use will be illustrated below using a concrete example.
For $U\in \mathcal{O}(X)$, consider the sheaf $C(U)$ of continuous real-valued functions $f: U \rightarrow \mathbb{R}$.
In Sh($X$), this {\em is\/} the real number object (which can be thought of as the set of variable real numbers).
Let $z$ denote a variable that can be assigned real-number values, and suppose we are given a formula $\phi(z)$ in one free variable $z$.
An example of such a formula is $(z>0)$. In {\bf Set\/} (or Sh($1$)), for any real number $\alpha$, a formula such as $(\alpha >0)$ evaluates
to {\bf True\/} or {\bf False\/}. If {\bf True\/}, it is possible to assert that $\alpha >0$, and an alternative way of writing this would be $\alpha \in \{z|z>0\}$.
For a general formula $\phi(z)$ in {\bf Set\/} one defines $\{z|\phi(z)\}$ as the inverse image of {\bf True\/} under $\phi$ (i.e., $\phi^{-1}({\bf True\/})$).
It is possible to mimic set-theoretic constructions like $\{z|\phi(z)\}$ in Sh($X$) as well.\footnote{In Sh($X$), assuming $\Omega$ denotes
the set of possible truth values, $\{z|\phi(z)\}$ is the pullback of the arrow $true: 1 \rightarrow \Omega$ along $\phi$.}
Continuing with the example where $\phi(z)$ denotes the formula $(z>0)$, for any $\alpha: U \rightarrow \mathbb{R}$, statements like $\alpha > 0$
will, in Sh($X$), have truth values that are open sets in $\mathcal{O}(U)$ (i.e., the open set $W$ such that $\alpha(x)>0$ for $x\in W$).
The statement $\alpha \in \{ z | \phi(z) \}$ will be taken to indicate that the truth value of $\phi(\alpha)$ is the largest possible open set (i.e., all of $U$).
So, when $\phi(z)$ denotes $(z>0)$, $\alpha \in \{ z | \phi(z) \}$ indicates that $\llbracket \alpha>0 \rrbracket = U$.
In Sh($X$), $\{z|\phi(z)\}$ is a {\em subsheaf\/} of $C$, which means (1)~$\{z|\phi(z)\}(V) \subseteq C(V)$ for all $V\subseteq U$,
and (2)~for all $f \in C(U)$ and open cover $\{U_i\}$ with $\bigcup_i U_i = U$ one has $f \in \{z|\phi(z)\}(U)$ if and only if  $f|_{U_i} \in \{z|\phi(z)\}(U_i)$.
The forcing relation $\Vdash$ is now defined as follows:
given an element $\alpha: U \rightarrow \mathbb{R}$, we have $U \Vdash \phi(\alpha)$ if and only if $\alpha \in \{z|\phi(z)\}(U)$.
The following two properties follow:
\begin{enumerate}
\item {\bf Monotonicity.\/} If $U\Vdash \phi(\alpha)$ then for open $V\subseteq U$, $V\Vdash \phi(\alpha|_V)$.
\item {\bf Local Character.\/} Give an open covering $\{U_i\}$ of $U$, if $U_i \Vdash \phi(\alpha|_{U_i})$ for all $i$,
then $U\Vdash \phi(\alpha)$.
\end{enumerate}
The following rules describe the forcing relation $\Vdash$ for sentences constructed using logical connectives and quantifiers (see Theorem~1, p.~304
and Theorem~1, p. 316 of \citep{maclane1996}). If we have two formulae in one free variable, $\phi(z)$ and $\psi(z)$, then the following apply:
\begin{enumerate}
\item $U \Vdash \phi(\alpha) \wedge \psi(\alpha)$ if and only if $U \Vdash \phi(\alpha)$ and $U \Vdash \psi(\alpha)$.
\item $U \Vdash \phi(\alpha) \vee \psi(\alpha)$ if and only if there is an open covering $\{U_i\}$ of $U$ such that for each $i$ we have
$U_i \Vdash \phi(\alpha|_{U_i})$ or $U_i \Vdash \psi(\alpha|_{U_i})$.
\item $U \Vdash \phi(\alpha) \Rightarrow \psi(\alpha)$ if and only if for all open $V \subseteq U$, $V \Vdash \phi(\alpha|_V)$ implies
$V \Vdash \psi(\alpha|_V)$.
\item $U \Vdash \neg\phi(\alpha)$ if and only if for no non-empty open set $V \subseteq U$ is it the case that $V \Vdash \phi(\alpha|_V)$.
\end{enumerate}
If $\phi(y,z)$ is a formula with two free variables, where $y$ is an element of a (possibly different) sheaf $G$ (e.g., $G(U) = \{f: U \rightarrow \mathbb{R}^2\}$),
\begin{enumerate}
\setcounter{enumi}{4}
\item $U \Vdash \exists y \phi(y,\alpha)$ if and only if there are a covering $\{U_i\}$ of $U$ and elements $\beta_i \in G(U_i)$ such that
$U_i \Vdash \phi(\beta_i,\alpha|_{U_i})$ for each $i$.
\item $U \Vdash \forall y \phi(y,\alpha)$ if and only if for all open $V\subseteq U$ and $\beta\in G(V)$ one has $V \Vdash \phi(\beta, \alpha|_V)$.
\end{enumerate}
These rules are used repeatedly in the paper. The rules for conjunction and existential quantifier are noteworthy because constructive principles
enter here. In the former case,
at every $x$ of $X$, there is a neighborhood $N_x$ over which one of $\phi(\alpha|_{N_x})$ or $\psi(\alpha|_{N_x})$ must hold. In other words,
one of $\phi(\alpha)$ or $\psi(\alpha)$ must be confirmed. In the latter case, the rule asserts that there is a neighborhood $N_x$ and a $\beta_x$ such that
$\phi(\beta_x,\alpha|_{N_x})$ holds. I.e., there is a witness $\beta$ to the existence of a $y$ such that $\phi(y, \alpha)$ holds. It is also worth noting the
local nature of the statements for conjunction and existence with the reference to open coverings. A number of properties of real and natural numbers in a sheaf
hold only in this local sense. Consider, for instance, the Archimedean property that, for every real number $\gamma$, there exists a natural number $n$ with $n>\gamma$.
Let $X = \mathbb{R}$ with the standard topology and consider the real number $\gamma(x)=x$ in Sh($X$). Since $X$ is connected, the natural numbers are the
constant functions with values in $\mathbb{N}$. Clearly, the Archimedean property cannot hold for global functions. However, it is possible to take an open covering
$\{U_i\}$ of $X$ and find a constant function greater than $\gamma$ over each $U_i$.

\end{document}